\documentclass[aps, prx, superscriptaddress, 10pt]{revtex4-2}

\usepackage{amsmath, amssymb}
\usepackage{braket}
\usepackage{bm}
\usepackage{amsthm}
\usepackage{todonotes}
\usepackage{ulem}
\usepackage[ruled, linesnumbered]{algorithm2e}

\newcommand{\pdif}[2]{\frac{\partial #1}{\partial #2}}
\newcommand{\pddif}[2]{\frac{\partial^2 #1}{\partial #2^2}}
\newcommand{\ketbra}[2]{\ket{#1} \! \bra{#2}}

\theoremstyle{plain}

\newtheorem*{thm*}{Theorem}

\newtheorem*{lem*}{Lemma}

\newtheorem*{cor*}{Corollary}
\theoremstyle{remark}

\newtheorem*{rem*}{Remark}

\begin{document}

\title{Quantum algorithm for partial differential equations of non-conservative systems \\with spatially varying parameters}

\author{Yuki Sato}
\email[]{yuki-sato@mosk.tytlabs.co.jp}
\affiliation{Toyota Central R\&D Labs., Inc., 1-4-14, Koraku, Bunkyo-ku, Tokyo, 112-0004, Japan}
\affiliation{Quantum Computing Center, Keio University, 3-14-1 Hiyoshi, Kohoku-ku, Yokohama, Kanagawa, 223-8522, Japan}

\author{Hiroyuki Tezuka}
\affiliation{Advanced Research Laboratory, Research Platform, Sony Group Corporation, 1-7-1 Konan, Minato-ku, Tokyo, 108-0075, Japan}
\affiliation{Quantum Computing Center, Keio University, 3-14-1 Hiyoshi, Kohoku-ku, Yokohama, Kanagawa, 223-8522, Japan}

\author{Ruho Kondo}
\affiliation{Toyota Central R\&D Labs., Inc., 1-4-14, Koraku, Bunkyo-ku, Tokyo, 112-0004, Japan}
\affiliation{Quantum Computing Center, Keio University, 3-14-1 Hiyoshi, Kohoku-ku, Yokohama, Kanagawa, 223-8522, Japan}

\author{Naoki Yamamoto}
\affiliation{Department of Applied Physics and Physico-Informatics, Keio University, Hiyoshi 3-14-1, Kohoku-ku, Yokohama, Kanagawa, 223-8522, Japan}
\affiliation{Quantum Computing Center, Keio University, 3-14-1 Hiyoshi, Kohoku-ku, Yokohama, Kanagawa, 223-8522, Japan}

\begin{abstract}
Partial differential equations (PDEs) are crucial for modeling various physical phenomena such as heat transfer, fluid flow, and electromagnetic waves.
In computer-aided engineering (CAE), the ability to handle fine resolutions and large computational models is essential for improving product performance and reducing development costs. 
However, solving large-scale PDEs, particularly for systems with spatially varying material properties, poses significant computational challenges. 
In this paper, we propose a quantum algorithm for solving second-order linear PDEs of non-conservative systems with spatially varying parameters, using the linear combination of Hamiltonian simulation (LCHS) method. 
Our approach transforms those PDEs into ordinary differential equations represented by qubit operators, through spatial discretization using the finite difference method. 
Then, we provide an algorithm that efficiently constructs the operator corresponding to the spatially varying parameters of PDEs via a logic minimization technique, which reduces the number of terms and subsequently the circuit depth. 
We also develop a scalable method for realizing a quantum circuit for LCHS, using a tensor-network-based technique, specifically a matrix product state (MPS). 
We validate our method with applications to the acoustic equation with spatially varying parameters and the dissipative heat equation. 
Our approach includes a detailed recipe for constructing quantum circuits for PDEs, leveraging efficient encoding of spatially varying parameters of PDEs and scalable implementation of LCHS, which we believe marks a significant step towards advancing quantum computing's role in solving practical engineering problems.
\end{abstract}

\maketitle

\section{Introduction}

Partial differential equations (PDEs) are fundamental mathematical models that describe the dynamics of various physical phenomena such as heat transfer, fluid flow, and electromagnetic waves~\cite{renardy2004introduction}. 
PDEs enable us to understand and predict the behavior of complex systems across a broad field of science and engineering. 
However, solving PDEs for extremely large-scale problems (i.e., problems having an extremely large number of degree of freedom) presents significant computational challenges, especially when dealing with systems having spatially varying material properties, that require high-resolution discretization to achieve accurate results.

As a practical use case of PDEs, computer-aided engineering~(CAE) has become indispensable in today's industries, being used in various fields such as the automotive industry, aerospace industry, architecture, and manufacturing.
The performance of CAE is directly linked to the competitiveness of products.
More concretely, fine resolution, large computational models, and rapid calculations contribute to the development of higher-performing products in a shorter period and at reduced developing costs. 
One clear example where relatively simple equations require substantial computational cost or memory is elastic wave simulation and acoustic simulation with a spatially varying parameter (i.e., density and speed of sound, respectively). 
The equations describing acoustics are linear PDEs, which are well-understood in computational mechanics contexts. 
However, when attempting to analyze the acoustics of large spaces like concert halls, i.e., room acoustics, precisely, a massive amount of memory and fine resolution are necessary~\cite{kuttruff2016room, prinn2023review}. 
In addition to that, the elastic wave simulation is used in geophysics for researching the structure of the Earth and analysing seismic wave propagation nature~\cite{10.1093/gji/ggac306}, and various techniques have been actively developed~\cite{virieux2009overview, 8931232/DeepLearning}.

Quantum computing emerges as a promising alternative that offers the potential to perform more efficient computation than classical computing, leveraging the principles of quantum mechanics.
There are several potential applications, including quantum phase estimation~\cite{nielsen2010quantum}, linear system solvers~\cite{harrow2009quantum, childs2017quantum, cao2013quantum}, and Hamiltonian simulation~\cite{lloyd1996universal, berry2015simulating, nielsen2010quantum, kim2023evidence, mc2023classically, loaiza2023reducing}.
Hamiltonian simulation is a technique to simulate a time evolution of a quantum system with the Hamiltonian $\mathcal{H}$; its aim is to implement the time evolution operator $\exp(-i \mathcal{H} \tau)$ with the time increment $\tau$ on a quantum computer.
The typical system Hamiltonian includes Ising Hamiltonians~\cite{kim2023evidence, mc2023classically} and molecular Hamiltonians~\cite{loaiza2023reducing}.
Several works also applied Hamiltonian simulation techniques for simulating the classical conservative systems by mapping the governing equations of classical systems to the Schr\"{o}dinger equation~\cite{costa2019quantum, babbush2023exponential, miyamoto2024quantum, wright2024noisy}.
Costa et al.~\cite{costa2019quantum} proposed a method of simulating the wave equation via Hamiltonian simulation where the system Hamiltonian is given as the incidence matrix of the discretized simulation space.
Babbush et al.~\cite{babbush2023exponential} extended this approach to propose a quantum simulation technique to simulate classical coupled oscillators with the rigorous proof of the exponential speedup over any classical algorithms.
Miyamoto et al.~\cite{miyamoto2024quantum} proposed a Hamiltonian-simulation-based quantum algorithm that solves the Vlasov equation for large-scale structure simulation.
Wright et al.~\cite{wright2024noisy} simulated one-dimensional wave equation using a real quantum device.
Recently, Schade et al.~\cite{10.1093/gji/ggae160} posed a quantum computing concept for simulating one-dimensional elastic wave propagation in spatially varying density and provide the potential of exponential speedup. 
Note that these approaches are limited to handling PDEs of conservative systems that can be mapped to the Schr\"{o}dinger equation; that is, the system with time evolution operator represented by a unitary operator.
Because most classical systems are non-conservative, it is important to develop a Hamiltonian-simulation-based approaches applicable to those practical systems.

Actually in the literature, we find several methods that can handle non-unitary operations on a quantum computer. 
The so-called Schr\"{o}dingerisation method~\cite{jin2023aquantum, jin2023bquantum} is an effective such approach for solving general ordinary differential equations (ODEs) by transforming them to the Schr\"{o}dinger equation via the warped phase transformation.
Another approach to deal with non-unitary opearations is the linear combination of unitaries (LCU)~\cite{childs2021theory, meister2022tailoring}, which has also been applied to the Hamiltonian simulation tasks~\cite{childs2012hamiltonian, berry2015simulating}.
Recently, based on the LCU, An et al.~\cite{an2023linear} proposed the so-called linear combination of Hamiltonian simulation (LCHS) method for solving linear ODEs with the guarantee of the optimal state preparation cost. 
While Ref.~\cite{an2023linear} mainly focuses on the theoretical discussion, it significantly showcases the potential for quantum computers to handle ODEs.
Since a PDE can be reduced to an ODE by discretization in the space direction, the PDE simulation boils down to an ODE one that can be implemented on a quantum device with the help of LCHS technique. 
However, PDEs considered in the practical scenario of CAE have spatially varying parameters as well as non-conservative properties, concrete and efficient encoding of which into the quantum algorithm has not been developed yet. 
Also, explicit implementation of LCHS on a quantum circuit involving oracles is still unclear.

This paper addresses the above two points. That is, we propose a quantum algorithm for solving linear PDEs of non-conservative systems with spatially varying parameters via LCHS. 
The contributions of this paper are summarized as follows. 
\begin{itemize}
    \item We provide a systematic method for transforming second-order linear PDEs to the ODEs represented as qubit operators, by the spatial discretization using the finite difference method.
    
    \item We construct the operator corresponding to the spatially varying parameters of PDEs in an efficient manner owing to a logic minimization technique. 
    This technique automatically generates the operator with the number of terms as small as possible and thus significantly reduces the subsequent circuit depth.
    
    \item We provide a tensor-network-based method for approximately constructing the quantum circuit for an oracle required for LCU. 
    Specifically, we derive a matrix product state (MPS) and encode it into a quantum circuit for the LCU, with the help of an optimization method. 
    This technique enables us to construct and implement the quantum circuit for LCU in a scalable and explicit manner, as both the tensor-network-based technique and the resulting quantum circuit are scalable.

    \item We demonstrate the validity of the proposed method by applying it to the acoustic equation with a spatially varying speed of sound, and moreover the dissipative heat equation which is a typical example of non-unitary dynamics.
\end{itemize}

\begin{figure}[t]
    \centering
    \includegraphics[width=\textwidth]{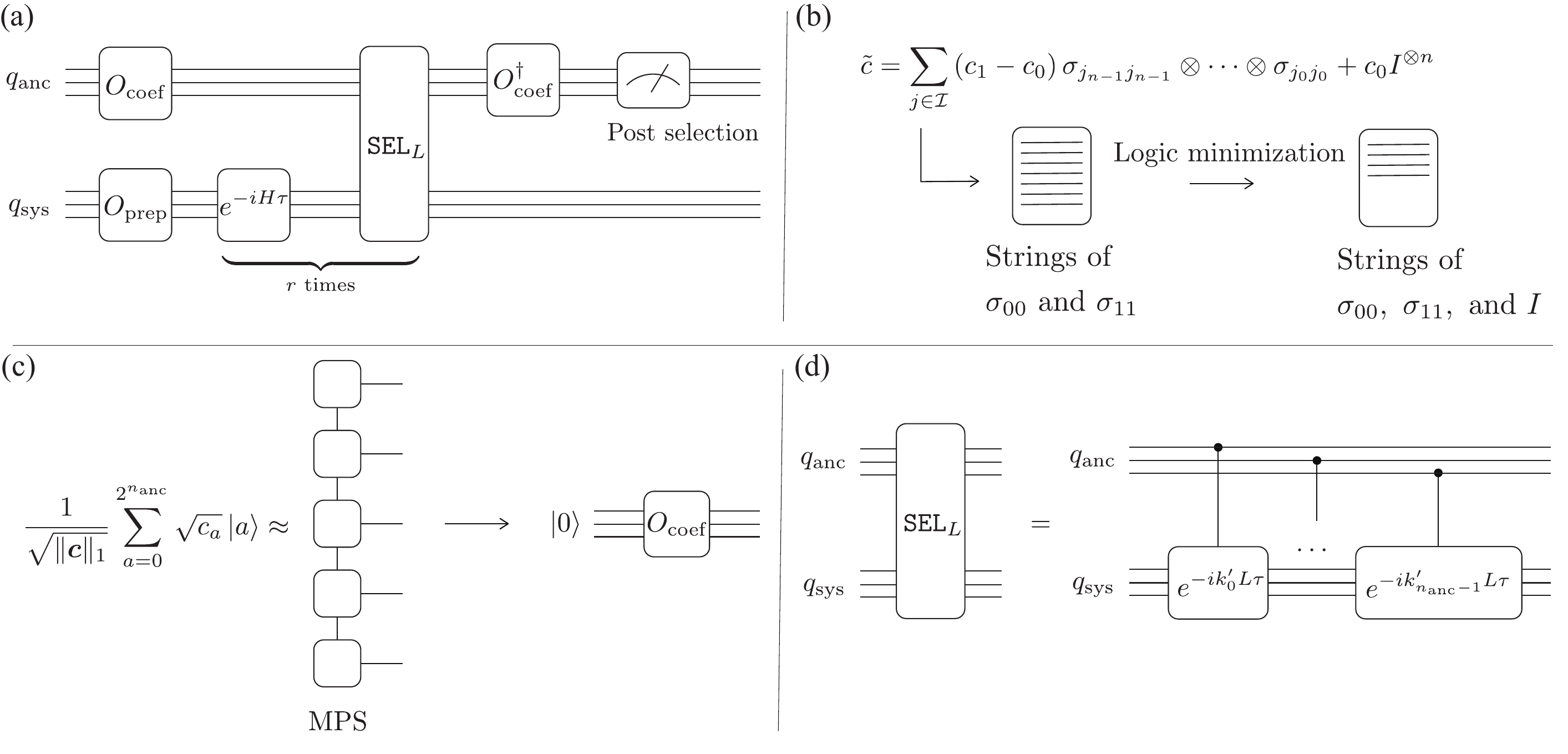}
    \caption{Overview of the proposed method. (a) Quantum circuit of LCHS based on the work by An et al.~\cite{an2023linear}. 
    (b) Efficient representation of Hamiltonian via logic minimizer techniques, which is described in Section~\ref{sec:discretization}. 
    (c) Implementation of the coefficient oracle for LCU, $O_\mathrm{coef}$, via the tensor network techniques, which is described in Section~\ref{sec:coefficient_oracle}. 
    (d) Quantum circuit of the select oracle $\texttt{SEL}_L$, which is described in Section~\ref{sec:procedure}.}
    \label{fig:overview}
\end{figure}

Figure~\ref{fig:overview} shows an overview of our proposed method, which will be described in detail with the concrete algorithm in Section~\ref{sec:procedure}. 
Our algorithm is based on LCHS, which will be reviewed in Section~\ref{sec:lchs}.
The quantum circuit for LCHS is summarized in Fig.~\ref{fig:overview}(a), which requires the state preparation oracle $O_\mathrm{prep}$, the coefficient oracle for the LCU $O_\mathrm{coef}$, the Hamiltonian simulation oracle $O_H(\tau)$, and the select oracle $\texttt{SEL}_L(\tau)$.
The state preparation oracle $O_\mathrm{prep}$ can be implemented naively for simple initial states, and even for general cases, methods for approximate implementation using tensor networks or parameterized quantum circuits are known \cite{shirakawa2024automatic, nakaji2022approximate, melnikov2023quantum}.
The Hamiltonian simulation oracle with finite difference operators can be efficiently implemented as described in the previous method~\cite{sato2024hamiltonian}. 
In the same manner, the select oracle $\texttt{SEL}_L(\tau)$ can be efficiently implemented.
Here, \textit{efficiently} means that they can be implemented with polynomial-depth quantum circuits. 
However, the previous method focused on PDEs with spatially uniform parameters.
Thus, the present study proposes the method for simulating PDEs of non-conservative systems with spatially-varying parameters, in Section~\ref{sec:discretization}.
In particular, we will describe how the Hamiltonian describing the spatially varying parameters can be constructed in an efficient manner using a logic minimization technique (Fig.~\ref{fig:overview}(b)).
Furthermore, we will consider the approximate implementation of the coefficient oracle for the LCU, $O_\mathrm{coef}$, based on a tensor network technique in Section~\ref{sec:coefficient_oracle} (Fig.~\ref{fig:overview}(c)).

Based on the notation used in Section.~\ref{sec:preliminary}, our algorithm requires $n_\mathrm{anc} \sim \mathcal{O}(\log (\| \bm{L} \|_2 T / \varepsilon^2))$ ancilla qubits where $\bm{L}$ is the Hermitian part of the coefficient matrix, $T$ is the simulation time, $\varepsilon$ is the precision parameter.
Owing to the tensor-network-based approximation, we can implement the coefficient oracle $O_\mathrm{coef}$ using $\mathcal{O}(n_\mathrm{anc})$ non-local gates.
Assuming that the Hamiltonian including spatially varying parameters is represented by at most $\mathrm{poly}(n)$ terms where the space is discretized into $2^n$ nodes, we can implement the Hamiltonian simulation oracle $O_H(\tau)$ using at most $\mathcal{O}(\mathrm{poly}(n) T^{3/2} / \varepsilon_t^{1/2})$ non-local gates~\cite{sato2024hamiltonian} where $\varepsilon_t$ is the precision parameter of the time integration.
Similarly, we can implement the select oracle $\texttt{SEL}_L(\tau)$ using at most $\mathcal{O}(n_\mathrm{anc} \mathrm{poly}(n) T^{3/2} / \varepsilon_t^{1/2})$ non-local gates.
Note that we herein assume that these oracles are implemented by the second-order Suzuki-Trotter decomposition while the state-of-the-art algorithms~\cite{low2019hamiltonian, martyn2021grand} achieved the time complexity $\mathcal{O}(\mathrm{poly}(\mathrm{log}(1/\varepsilon_t)))$.
The expected number of repetition to measure the outcome of all zeros in the ancilla qubits, which is success, is of the order $\mathcal{O}(\| \bm{w}(0) \| / \| \bm{w}(T) \|)$, which is the ratio of the norm of the state vector at the initial and final times.

\section{Preliminaries} \label{sec:preliminary}

Section~\ref{sec:lchs} briefly reviews the method of linear combination of Hamiltonian simulation (LCHS)~\cite{an2023linear} for representing a non-conservative dynamics in non-unitary quantum mechanical framework. 
Section~\ref{sec:diff_mpo} introduces the finite difference operators which will be used for mapping a classical PDE to a quantum non-unitary dynamics.

\subsection{Linear combination of Hamiltonian simulation} \label{sec:lchs}

Let $\bm{w}(t) \in \mathbb{C}^{N}$ be an $N$-dimensional vector of variables, $\bm{A}(t) \in \mathbb{C}^{N \times N}$ the time-dependent coefficient matrix, and $\bm{b}(t) \in \mathbb{C}^{N}$ the inhomogeneous term that form a (classical) linear ordinary differential equation (ODE):
\begin{align} \label{eq:ode}
    \frac{\mathrm{d} \bm{w}(t)}{\mathrm{d}t} &= - \bm{A}(t) \bm{w}(t) + \bm{b}(t).
\end{align}
The LCHS gives an analytic solution for this ODE, as follows:
\begin{align}
    \bm{w}(t) &= \int_\mathbb{R} \frac{1}{\pi (1 + k^2)} \mathcal{T} \exp \left( - i \int_0^t \left( \bm{H}(\tau) + k\bm{L}(\tau) \right) \mathrm{d}\tau \right) \bm{w}(0) \mathrm{d}k \nonumber \\
    &+ \int_0^t \int_\mathbb{R} \frac{1}{\pi (1 + k^2)} \mathcal{T} \exp \left( - i \int_\tau^t \left( \bm{H}(\tau') + k\bm{L}(\tau') \right) \mathrm{d}\tau' \right) \bm{b}(\tau) \mathrm{d}k \mathrm{d}\tau, \label{eq:lchs}
\end{align}
where $\mathcal{T}$ is the time-ordering operator. 
$\bm{L}(t):=(\bm{A}(t) + \bm{A}^\dagger (t))/2$ and $\bm{H}(t):=(\bm{A}(t) - \bm{A}^\dagger (t))/2i$ are Hermitian operators, which recover $\bm{A}(t) = \bm{L}(t) + i \bm{H}(t)$ where $\bm{L}(t)$ must be positive semidefinite.
This condition can be always satisfied without loss of generality by changing the variable $\bm{w}(t)$ to $\bm{w}'(t):=e^{-at} \bm{w}(t)$ for a constant $a$~\cite{an2023linear}.
Equation~\eqref{eq:lchs} represents the integration of 
the unitary time evolution driven by the Hamiltonian $\bm{H}(\tau) + k\bm{L}(\tau)$ with respect to $k$, and thus it is called the linear combination of Hamiltonian simulation, which is a special form of LCU. 
For ODEs with the time-independent coefficient matrix $\bm{A} := \bm{A}(t)$ and $\bm{b}(t)=0$, the formula reads
\begin{align}
    \bm{w}(t) &= \int_\mathbb{R} \frac{1}{\pi (1 + k^2)} \exp \left( - i \left( \bm{H} + k\bm{L} \right) t \right) \bm{w}(0) \mathrm{d}k, \label{eq:lchs_time_independent}
\end{align}
where $\bm{L} := (\bm{A} + \bm{A}^\dagger )/2$ and $\bm{H} := (\bm{A} - \bm{A}^\dagger )/2i$.
When furthermore $\bm{L}=0$, the ODE~\eqref{eq:ode} corresponds to the Schr\"{o}dinger equation $\mathrm{d} \bm{w}(t) / \mathrm{d} t = - i \bm{H} \bm{w}(t)$ and the solution~\eqref{eq:lchs} coincides with that of Hamiltonian simulation:
\begin{align}
    \bm{w}(t) &= \exp \left( - i \bm{H}t \right) \bm{w}(0).
\end{align}

In this study, we focus on PDEs with time-independent but spatially varying parameters, which results in the form of Eq.~\eqref{eq:lchs_time_independent}.
The procedure for implementing Eq.~\eqref{eq:lchs_time_independent} on a quantum computer is summarized in Fig.~\ref{fig:overview}(a) where each operation is described as follows:
\begin{enumerate}
    \item Prepare $n$ qubits (referred to as the system register) to embed the solution $\bm{w}(t)$ of the ODE and $n_\mathrm{anc}$ ancilla qubits (referred to as the ancilla register) for the LCU. 
    The number of necessary ancilla qubits depends on discretization of $k$, that is, the number of integration points $M$ in the numerical integration with respect to $k$, in the form $n_\mathrm{anc} = \lceil \log_2 M \rceil$. 
    With the integration points $\{ k_a \}_{a=0}^{M-1}$, we let $c_a$ denote the coefficient of LCU, i.e., $c_a := \omega_a / \pi (1 + k_a^2)$ where $\omega_a$ is the weight for numerical integration, and approximate Eq.~\eqref{eq:lchs_time_independent} by $\bm{w}(t) \approx \sum_{a=0}^{M-1} c_a \exp \left( - i \left( \bm{H} + k_a \bm{L} \right) t \right) \bm{w}(0)$. Since $U_a := \exp \left( - i \left( \bm{H} + k_a \bm{L} \right) t \right)$ is a unitary, this is exactly LCU of $\sum_{a=0}^{M-1} c_a U_a$.
    \item Initialize the qubits to $\ket{0}^{\otimes n_\mathrm{anc} } \otimes \ket{0}^{\otimes n }$.
    \item Apply the state preparation oracle $O_\mathrm{prep}: \ket{0}^{\otimes n} \rightarrow \ket{w(0)} := \bm{w}(0) / \| \bm{w}(0) \|$ to the system register to prepare the quantum state $\ket{w(0)}$ representing the normalized initial state of $\bm{w}(0)$.
    \item Apply the coefficient oracle for the LCU, $O_\mathrm{coef}: \ket{0} \rightarrow (1 / \sqrt{\| \bm{c} \|_1})\sum_{a=0}^{M-1} \sqrt{c_a} \ket{a}$, to the ancilla register.
    \item With the Hamiltonian simulation oracle $O_H(\tau):=e^{-i \bm{H} \tau}$ and the select oracle $\texttt{SEL}_L(\tau):= \sum_{a=0}^{M-1} \ket{a} \! \bra{a} \otimes e^{-i k_a \bm{L} \tau}$, alternatively apply $O_H$ to the system register and $\texttt{SEL}_L$ to the whole system $r$ times, where $r := T / \tau$ is the number of steps for time evolution.
    \item Apply the inverse of the coefficient oracle, $O_\mathrm{coef}^\dagger$, to the ancilla register.
\end{enumerate}
Following this procedure, we obtain the output state of whole system before the measurement as 
$(1 / \| \bm{c} \|_1 )\ket{0}^{\otimes n_\mathrm{anc}} \otimes \exp \left( - \bm{A} t \right) \ket{w(0)} + \ket{\perp}$, where $\ket{\perp}$ represents a state orthogonal to the first term. 
Then, 
when the measurement result of the ancilla register is all zeros, the system register approximates the state \eqref{eq:lchs_time_independent} in the sense of numerical integration (i.e., approximation via quadrature by parts and Trotterization). 
The expected number of repetition to successfully get the measurement outcome of all zeros in the ancilla register is 
of the order $\mathcal{O}(\| \bm{c} \|_1^2 \| \bm{w}(0) \|^2 / \| \bm{w}(T) \|^2)$, but this can be improved to $\mathcal{O}(\| \bm{c} \|_1 \| \bm{w}(0) \| / \| \bm{w}(T) \|)$ using the amplitude amplification technique (note that $\|\bm{w}(0)\| \geq \|\bm{w}(T)\|$ in general).

We aware that the truncation error of the numerical integration of $k$ has been improved by using exponentially decaying kernel function~\cite{an2023quantum}, but we employ the original version~\cite{an2023linear} for ease to implement by LCU. 
Note that the authors of LCHS also proposed a hybrid quantum-classical implementation of LCU when ones are only interested in obtaining an expected value of an observable $\bm{w}(t)^\dagger O \bm{w}(t)$~\cite{an2023linear}.
In this case, we no longer have to implement $O_\mathrm{coef}$ to the ancilla register but we need classical Monte Carlo sampling and one ancilla qubit.

\subsection{Finite difference operators and its representation in qubits} \label{sec:diff_mpo}

Let us consider a $d$-dimensional domain $\Omega \subset \mathbb{R}^d$ with an orthogonal coordinate $\bm{x}:=(x_0, \dots x_{d-1})$ and a scalar field $u(\bm{x})$ defined on the domain.
Discretizing the domain $\Omega$ into a grid consisting of $N=\prod_{\mu=0}^{d-1} N_\mu$ nodes, we represent the scalar field $u(\bm{x})$ as the value at each node, $\bm{u}:=[u(\bm{x}^{[0]}), u(\bm{x}^{[1]}), \dots, u(\bm{x}^{[N-1]})]^\top$, where $\bm{x}^{[j]}$ is the coordinate of the $j$-th node.
The finite difference method (FDM) approximates the spatial gradient of the scalar field using its node values $\bm{u}$.
For instance, the forward difference scheme approximates the first-order spatial derivative as
\begin{align} \label{eq:forward_diff_operator}
    \pdif{u(\bm{x}^{[j]})}{x_\mu} = \frac{u_{j_{(\mu, 1)}} - u_j}{ h } + O(h),
\end{align}
where $u_{j_{(\mu, 1)}} := u(\bm{x}^{[j]} + h \bm{e}_\mu)$ with the orthonormal basis $\bm{e}_\mu$ along with the $x_\mu$-axis, $u_j := u(\bm{x}^{[j]})$ and $h$ is the interval of adjacent nodes.
That is, $\bm{x}^{[j]} + h \bm{e}_\mu$ corresponds to the coordinate of the $j_{(\mu, 1)}$-th node which locates next to the $j$-th node along with the $x_\mu$-axis.
Similarly, the central difference scheme with the fourth-order accuracy is given as
\begin{align}
    \pdif{u(\bm{x}^{[j]})}{x_\mu} = \frac{-u_{j_{(\mu, 2)}} + 8 u_{j_{(\mu, 1)}} - 8u_{j_{(\mu, -1)}} + u_{j_{(\mu, -2)}}}{ 12h } + O(h^4),
\end{align}
where $u_{j_{(\mu,2)}} := u(\bm{x}^{[j]} + 2h \bm{e}_\mu)$, $u_{j_{(\mu, -1)}} := u(\bm{x}^{[j]} - h \bm{e}_\mu)$ and $u_{j_{(\mu, -2)}} := u(\bm{x}^{[j]} - 2h \bm{e}_\mu)$.
In the previous research~\cite{sato2024hamiltonian}, we showed that the finite difference operators including forward, backward and central differencing schemes, can be represented using operators for qubit systems, which we will briefly review here.

To represent the finite difference operator in terms of qubit operators, we first map the discretized scalar field $\bm{u}$ into the system state, with the assumption that $N_\mu=2^{n_\mu}$, as
\begin{align} \label{eq:ket_u}
    \ket{u}:= \frac{1}{\| \bm{u} \|} \sum_{j=0}^{2^n-1} u(\bm{x}^{[j]}) \ket{j},
\end{align}
where $n := \sum_{\mu=0}^{d-1} n_\mu$, $\ket{j}:=\ket{j_{n-1} j_{n-2} \dots j_0}$ with $j_{n-1}, j_{n-2}, \dots, j_0 \in \{0, 1\}$ is the computational basis.
Let $\sigma_{01}$ and $\sigma_{10}$ denote the ladder operators defined as 
\begin{equation}
	\sigma_{01} := \begin{pmatrix}
		0 & 1 \\
		0 & 0
	\end{pmatrix}, ~ \sigma_{10} := \begin{pmatrix}
		0 & 0 \\
		1 & 0
	\end{pmatrix}.
\end{equation}
Then, the shift operators $S^{-}(n_\mu) := \sum_{j'=1}^{2^{n_\mu}-1} \ket{j'-1} \! \bra{j'}$ and $S^{+}(n_\mu) := \sum_{j'=1}^{2^{n_\mu}-1} \ket{j'} \! \bra{j'-1}$ acting on the system are represented as
\begin{align}
    S^{-}(n_\mu) &= \sum_{j'=1}^{n} I^{\otimes (n_\mu-j')} \otimes \sigma_{01} \otimes \sigma_{10}^{\otimes (j'-1)} \nonumber \\
        &= \sum_{j'=1}^{n_\mu} s^{-}(n_\mu, j'), \label{eq:shift_minus} \\
    S^{+}(n_\mu) &= \left( S^{-}(n_\mu) \right)^\dagger \nonumber \\
    &= \sum_{j'=1}^{n_\mu} I^{\otimes (n_\mu-j')} \otimes \sigma_{10} \otimes \sigma_{01}^{\otimes (j'-1)} \nonumber \\
    &= \sum_{j'=1}^{n_\mu} s^{+}(n_\mu, j'), \label{eq:shift_plus}
\end{align}
where
\begin{align}
    s^{-}(n_\mu, j') &:= I^{\otimes (n_\mu-j')} \otimes \sigma_{01} \otimes \sigma_{10}^{\otimes (j'-1)},  \\
    s^{+}(n_\mu, j') &:= I^{\otimes (n_\mu-j')} \otimes \sigma_{10} \otimes \sigma_{01}^{\otimes (j'-1)}.
\end{align}
Note that $\sigma_{ij}^{\otimes 0}$ is regarded as a scalar 1. 
For instance, in the case of $n_\mu=2$, we have $S^{-}(2) = \ket{0} \! \bra{1} + \ket{1} \! \bra{2} + \ket{2} \! \bra{3} = \sigma_{01} \otimes \sigma_{10} + I \otimes \sigma_{01}$. 
Using the shift operators, we can represent various finite difference operators using qubit operators~\cite{sato2024hamiltonian}.
Actually, we obtain the following relationship:
\begin{align}
    \frac{1}{h} \left( S^{-}(n_\mu; \{ n_\mu \}) - I^{\otimes n} \right) \sum_{j=0}^{2^n-1} u(\bm{x}^{[j]}) \ket{j} = \sum_{j=0}^{2^n-1} \frac{ u_{j_{(\mu,1)}} - u_j }{ h } \ket{j},
\end{align}
where $S^{-}(n_\mu; \{ n_\mu \}) := I^{\otimes \sum_{\mu'=\mu+1}^{d-1} n_{\mu'}} \otimes S^{-}(n_\mu) \otimes I^{\otimes \sum_{\mu'=0}^{\mu-1} n_{\mu'}}$, and $u_{j_{(\mu,1)}} = 0$ when $\bm{x}^{[j]} + h \bm{e}_\mu$ is out of the domain $\Omega$, which exactly corresponds to the forward difference operator with the Dirichlet boundary condition. 
Similarly, we obtain the following relationship for the central difference scheme with the fourth-order accuracy:
\begin{align}
    &\frac{1}{12h} \left( - S^{-}(n_\mu; \{ n_\mu \})^2 + 8 S^{-}(n_\mu; \{ n_\mu \}) - 8 S^{+}(n_\mu; \{ n_\mu \}) + S^{+}(n_\mu; \{ n_\mu \})^2 \right) \sum_{j=0}^{2^n-1} u(\bm{x}^{[j]}) \ket{j} \nonumber \\
    &= \sum_{j=0}^{2^n-1} \frac{- u_{j_{(\mu, 2)}} + 8 u_{j_{(\mu, 1)}} - 8u_{j_{(\mu, -1)}} + u_{j_{(\mu, -2)}}}{ 12h } \ket{j},
\end{align}
where $u_{j_{(\mu, 2)}}$, $u_{j_{(\mu, 1)}}$, $u_{j_{(\mu, -1)}}$ and $u_{j_{(\mu, -2)}}$ are zero, respectively when the nodes on $\bm{x}^{[j]} + 2h \bm{e}_\mu$, $\bm{x}^{[j]} + h \bm{e}_\mu$, $\bm{x}^{[j]} - h \bm{e}_\mu$, and $\bm{x}^{[j]} - 2h \bm{e}_\mu$ are out of the domain $\Omega$, which exactly corresponds to the central difference operator with the Dirichlet boundary condition.

\section{Method} \label{sec:method}

\subsection{Problem of interest}

In this study, we focus on the following second-order and first-order linear PDEs in time defined on an open bounded set $\Omega \subset \mathbb{R}^d$ with the spatial coordinate $\bm{x}$:
\begin{align} \label{eq:pde}
    \varrho(\bm{x}) \pddif{u(t, \bm{x})}{t} + \zeta(\bm{x}) \pdif{u(t, \bm{x})}{t} - \nabla \cdot \kappa(\bm{x}) \nabla u(t, \bm{x}) + \alpha(\bm{x}) u(t, \bm{x}) = 0, \text{ for } (t, \bm{x}) \in (0, T] \times \Omega,
\end{align}
and
\begin{align} \label{eq:pde_1}
    \pdif{u(t, \bm{x})}{t} - \nabla \cdot \kappa(\bm{x}) \nabla u(t, \bm{x}) + \bm{\beta}(\bm{x}) \cdot \nabla u(t, \bm{x}) + \alpha(\bm{x}) u(t, \bm{x}) = 0, \text{ for } (t, \bm{x}) \in (0, T] \times \Omega,
\end{align}
under an appropriate initial condition. 
$u(t, \bm{x})$ is the scalar field. 
Also, $\varrho(\bm{x}) > 0$, $\zeta(\bm{x}) \geq 0$, $\kappa(\bm{x}) \geq 0$, $\bm{\beta}(\bm{x})$, and $\alpha(\bm{x}) \geq 0$ are spatially varying parameters that physically correspond to the density, the damping factor, the diffusion coefficient, the velocity, and the absorption coefficient, respectively. 
We consider the following boundary conditions for the PDE:
\begin{align} \label{eq:bc}
    \begin{cases}
        \bm{n}(\bm{x}) \cdot \kappa(\bm{x}) \nabla u(t, \bm{x}) = 0 & \text{for } (t, \bm{x}) \in (0, T] \times \Gamma_\mathrm{N} \\
        u(t, \bm{x}) = 0 & \text{for } (t, \bm{x}) \in (0, T] \times \Gamma_\mathrm{D},
    \end{cases}
\end{align}
where $\bm{n}(\bm{x})$ is the outward-pointing normal vector and $\Gamma_\mathrm{D} \cup \Gamma_\mathrm{N} = \partial \Omega$ with $\Gamma_\mathrm{D} \cap \Gamma_\mathrm{N}=\emptyset$ and the boundary of the domain, $\partial \Omega$.
The PDE \eqref{eq:pde} can be represented in the following form:
\begin{align} \label{eq:transformed_pde}
    \frac{\partial }{\partial t} \begin{pmatrix}
        \sqrt{\varrho} \dot{u} \vspace{1mm} \\
        \sqrt{\kappa} \nabla u \vspace{1mm} \\
        \sqrt{\alpha} u
    \end{pmatrix} &= - \begin{pmatrix}
        \frac{\zeta(\bm{x})}{\varrho(\bm{x})} & -\frac{1}{\sqrt{\varrho}(\bm{x})} \nabla^\top \sqrt{\kappa(\bm{x})} & \sqrt{\frac{\alpha(\bm{x})}{\varrho(\bm{x})}}  \vspace{1mm} \\
        -\sqrt{\kappa(\bm{x})} \nabla \frac{1}{\sqrt{\varrho(\bm{x})}} & 0 & 0 \vspace{1mm} \\
        - \sqrt{\frac{\alpha(\bm{x})}{\varrho(\bm{x})}} & 0 & 0
    \end{pmatrix} \begin{pmatrix}
        \sqrt{\varrho} \dot{u} \vspace{1mm} \\
        \sqrt{\kappa} \nabla u\vspace{1mm} \\
        \sqrt{\alpha} u
    \end{pmatrix}, \text{ for } (t, \bm{x}) \in (0, T] \times \Omega,
\end{align}
where $\dot{u} := \partial u(t, \bm{x}) / \partial t$. 
Note that we apply the operators in order from the right side; for instance, the product of $(2, 1)$-component of the matrix and the first component of the vector on the right-hand side is calculated as $\sqrt{\kappa} \nabla ((1 / \sqrt{\varrho}) \sqrt{\varrho} \dot{u}) = \sqrt{\kappa} \nabla \dot{u}$.
We omit dependence of variables on $(t, \bm{x})$ for simplicity. 
Although there are various mappings from the second-order PDE~\eqref{eq:pde} to a first-order one, Eq.~\eqref{eq:transformed_pde} has the following advantage. 
The key fact is that, as the dissipative nature of the target system, the norm of state vector of such equation must be reduced over time, and the reduction rate (will be shown to be $\| \bm{w}(0) \| / \| \bm{w}(T) \|$) directly represents the computational complexity for Hamiltonian simulation method; yet the reduction rate in the case of Eq.~\eqref{eq:transformed_pde} is the minimum in a certain sense as proven in Appendix~\ref{sec:mapping_pde}. 
Specifically, the norm reduction is governed only by the damping factor $\zeta(\bm{x})$.
For example, the acoustic wave equation we will focus on later in Eq.~\eqref{eq:acoustic} falls into Eq.~\eqref{eq:pde} by setting $\zeta(\bm{x}) = 0$, $\kappa(\bm{x}) = 1$, $\alpha(\bm{x}) = 0$, and $\rho(\bm{x}) = 1 / c(\bm{x})^2$ with $c(\bm{x})$ the spatially varying speed of sound.
At that moment, the boundary condition $\bm{n}(\bm{x}) \cdot \nabla u(t, \bm{x}) = 0$ corresponds to the sound hard boundary condition, i.e., the wall boundary, while $u(t, \bm{x}) = 0$ corresponds to the sound soft boundary condition, i.e., the free end.
Since the acoustic equation has no damping factor $(\zeta(\bm{x})=0)$, it can be encoded in Eq.~\eqref{eq:transformed_pde} without any norm reduction.
The PDE~\eqref{eq:pde_1} can be simply represented as
\begin{align} \label{eq:transformed_pde_1}
    \frac{\partial u}{\partial t} &= \left( \nabla \cdot \kappa \nabla  - \bm{\beta} \cdot \nabla - \alpha \right) u, \text{ for } (t, \bm{x}) \in (0, T] \times \Omega.
\end{align}
The reduction rate of the norm, which directly affects the computational complexity, is evaluated in Appendix~\ref{sec:norm_pde_1}. 
Specifically, the norm reduction is governed by the diffusion and absorption coefficients $\kappa(\bm{x})$ and $\alpha(\bm{x})$ under the conservation condition of $\bm{\beta}(\bm{x})$.
For example, the heat equation we will focus on later in Eq.~\eqref{eq:heat} falls into Eq.~\eqref{eq:pde_1} by setting $\beta(\bm{x}) = 0$ and $\alpha(\bm{x}) = 0$.
At that moment, the boundary condition $\bm{n}(\bm{x}) \cdot \kappa(\bm{x}) \nabla u(t, \bm{x}) = 0$ corresponds to the thermal insulation, while $u(t, \bm{x}) = 0$ corresponds to the prescribed temperature.
Since the heat equation has the diffusion coefficient $(\kappa(\bm{x}) > 0)$, it results in the norm reduction depending on the magnitude of the coefficient.
In the subsequent sections, we discuss how to spatially discretize Eqs.~\eqref{eq:transformed_pde} and \eqref{eq:transformed_pde_1} followed by representing them as qubit systems.

\subsection{Quantization of partial differential equation} \label{sec:discretization}

\subsubsection{State vector and spatial difference operator}

Consider to discretize Eq.~\eqref{eq:transformed_pde} in the space direction, which results in an ODE.
To this end, we discretize the domain $\Omega \subset \mathbb{R}^d$ into the lattice with $2^n$ nodes, of which the position of the $j$-th node is denoted as $\bm{x}^{[j]}$.
We assume the number of nodes along with the $x_\mu$-axis is $2^{n_\mu}$, i.e., $n = \sum_{\mu=0}^{d-1} n_\mu$ and the interval between the nodes is $h$.
Then, we associate the computational basis $\ket{j} = \ket{j^{[d-1]}} \dots \ket{j^{[0]}}$ of an $n$-qubit system with the $j$-th node, where $j^{[\mu]}$ counts the node along with the $x_\mu$-axis.
We now represent the vector $\bm{w}(t)$, accociated with the $n+ \lceil \log_2 (d+2) \rceil$-qubit system as follows:
\begin{align} \label{eq:w_vector}
    \bm{w}(t) = \sum_{j=0}^{2^n-1} \Bigg(
    \underbrace{w_{0, j}(t) \ket{0} \! \ket{j}\vphantom{\sum_{\mu=0}^{d-1} w_{\mu+1, j}(t) \ket{\mu + 1} \! \ket{j}}}_{\sqrt{\varrho}\dot{u}}
    + \underbrace{\sum_{\mu=0}^{d-1} w_{\mu+1, j}(t) \ket{\mu + 1} \! \ket{j}}_{\sqrt{\kappa}\nabla u}
    + \underbrace{w_{d+1, j}(t) \ket{d+1} \! \ket{j}\vphantom{\left(\sum_{\mu=0}^{d-1} w_{\mu+1, j}(t) \ket{\mu + 1} \! \ket{j}\right)}}_{\sqrt{\alpha}u}
    \Bigg),
\end{align}
where $w_{\cdot, \cdot}$ is a scalar variable depending on time $t$.
This vector $\bm{w}(t)$ will contain $\sqrt{\varrho}\dot{u}$, $\sqrt{\kappa} \nabla u$, and $\sqrt{\alpha} u$ as we describe later. 
Note that $\bm{w}(t)$ is not normalized; hence, we prefer to use the bold symbol $\bm{w}$ rather than the ket symbol because we consistently use the ket symbol to represent quantum states, i.e., vectors with the unit norm.

Next, let us discretize the matrix in Eq.~\eqref{eq:transformed_pde}.
Note first that we can discretize any spatially varying parameter $c(\bm{x})$ as a diagonal operator $\tilde{c}$, as follows:
\begin{align} \label{eq:c_diag}
    \tilde{c} = \sum_{j=0}^{2^n - 1} c(\bm{x}^{[j]}) \ket{j} \! \bra{j}.
\end{align}
Actually, letting $\bm{u}:=\sum_{j=0}^{2^n - 1} u(\bm{x}^{[j]}) \ket{j}$ be a scalar field $u(\bm{x})$ discretized on the lattice, the diagonal operator $\tilde{c}$ acts on $\bm{u}$ as
\begin{align}
    \tilde{c} \bm{u} = \sum_{j=0}^{2^n - 1} c(\bm{x}^{[j]}) u(\bm{x}^{[j]}) \ket{j},
\end{align}
which represents the product $c(\bm{x}) u(\bm{x})$ on the lattice.
In this way, we add the tilde to denote the diagonal operators 
corresponding to spatially varying parameters.
Discretizing the spatial gradient $\partial / \partial x_\mu$ by forward and backward difference operators, respectively denoted by $D^{+}_\mu$ and $D^{-}_\mu$, we define the following matrix $\bm{A}$, which is the discretized version of the matrix in Eq.~\eqref{eq:transformed_pde}:
\begin{align} \label{eq:A_matrix}
    \bm{A} &:= \ketbra{0}{0} \otimes \tilde{\varrho}^{-1} \tilde{\zeta} - \sum_{\mu=0}^{d-1} \ketbra{0}{\mu + 1} \otimes \tilde{\varrho}^{-\frac{1}{2}} D_\mu^{+} \tilde{\kappa}^{\frac{1}{2}} + \ketbra{0}{d+1} \otimes \tilde{\varrho}^{-\frac{1}{2}} \tilde{\alpha}^{\frac{1}{2}} \nonumber \\
    & \quad - \sum_{\mu=0}^{d-1} \ketbra{\mu + 1}{0} \otimes \tilde{\kappa}^{\frac{1}{2}} D_\mu^{-} \tilde{\varrho}^{-\frac{1}{2}} - \ketbra{d+1}{0} \otimes \tilde{\varrho}^{-\frac{1}{2}} \tilde{\alpha}^{\frac{1}{2}} \nonumber \\
    & \quad + \frac{1}{h} \sum_{\mu=0}^{d-1} \chi_\mathrm{D} (B_\mu^{+}) \ketbra{0}{\mu + 1} \otimes \tilde{\varrho}^{-\frac{1}{2}} \left( I^{\otimes \sum_{\mu'=\mu+1}^{d-1}} \otimes \sigma_{11}^{\otimes n_\mu - 1} \otimes \left( 2\sigma_{11} - \sigma_{10} \right) \otimes I^{\otimes \sum_{\mu'=0}^{\mu-1} n_{\mu'}} \right) \tilde{\kappa}^{\frac{1}{2}} \nonumber \\
    & \quad - \frac{1}{h} \sum_{\mu=0}^{d-1} \chi_\mathrm{N} (B_\mu^{-}) \ketbra{\mu + 1}{0} \otimes \tilde{\kappa}^{\frac{1}{2}} \left( I^{\otimes \sum_{\mu'=\mu+1}^{d-1}} \otimes \sigma_{00}^{\otimes n_\mu - 1} \otimes \left( 2\sigma_{00} - \sigma_{01} \right) \otimes I^{\otimes \sum_{\mu'=0}^{\mu-1} n_{\mu'}} \right) \tilde{\varrho}^{-\frac{1}{2}},
\end{align}
where $\sigma_{00} := \ketbra{0}{0}$, $\sigma_{11} := \ketbra{1}{1}$. 
Note again that for instance $\tilde{\varrho}$ represents a diagonal operator corresponding to $\varrho(\bm{x})$. 
Also, $B_\mu^{+}$ and $B_\mu^{-}$ denote the boundaries whose outward-pointing normal vectors are $\bm{e}_{\mu}$ and $-\bm{e}_{\mu}$, respectively. 
$\chi_\mathrm{N}$ and $\chi_\mathrm{D}$ are indicator functions defined as
\begin{align}
    \chi_\mathrm{N}(B_\mu^{\cdot}) &= \begin{cases}
        1 & \text{ if } B_\mu^{\cdot} \subset \Gamma_\mathrm{N} \\
        0 & \text{ otherwise }
    \end{cases}, \\
    \chi_\mathrm{D}(B_\mu^{\cdot}) &= \begin{cases}
        1 & \text{ if } B_\mu^{\cdot} \subset \Gamma_\mathrm{D} \\
        0 & \text{ otherwise },
    \end{cases}
\end{align}
respectively.
The third and fourth lines of Eq.~\eqref{eq:A_matrix} 
appear due to the boundary condition.
Specifically, the third line replaces the forward difference operator with the backward one at the boundary $B_\mu^{+}$ when the Dirichlet boundary condition is imposed on the boundary.
The fourth line replaces the backward difference operator with the forward one at the boundary $B_\mu^{-}$ when the Neumann boundary condition is imposed on the boundary.
The discretization in Eq.~\eqref{eq:A_matrix} is just one example and we can use another finite difference scheme made from the shift operators in Eqs.~\eqref{eq:shift_minus} and \eqref{eq:shift_plus}.
We can now consider the ODE $\mathrm{d} \bm{w} / \mathrm{d} t = - \bm{A} \bm{w}$ under the initial condition $w_{0, j}(0) = \sqrt{\varrho(\bm{x}^{[j]})}\dot{u}(0, \bm{x}^{[j]})$, $w_{\mu + 1, j}(0) = \sqrt{\kappa (\bm{x}^{[j]})} \partial u(0, \bm{x}^{[j]}) / \partial x_\mu$, and $w_{d + 1, j}(0) = \sqrt{\alpha (\bm{x}^{[j]})} u(0, \bm{x}^{[j]})$, which results in the solution of $w_{0, j}(t) = \sqrt{\varrho(\bm{x}^{[j]})}\dot{u}(t, \bm{x}^{[j]})$, $w_{\mu + 1, j}(t) = \sqrt{\kappa (\bm{x}^{[j]})} \partial u(t, \bm{x}^{[j]}) / \partial x_\mu$, and $w_{d + 1, j}(t) = \sqrt{\alpha (\bm{x}^{[j]})} u(t, \bm{x}^{[j]})$.

Next, we consider to discretize Eq.~\eqref{eq:transformed_pde_1} in the space direction under the same discretization of $\Omega$, representing the vector $\bm{w}(t)$, accociated with the qubit system as follows:
\begin{align} \label{eq:w_vector_1}
    \bm{w}(t) = \sum_{j=0}^{2^n-1} w_{j}(t) \ket{j}.
\end{align}
Accordingly, we discretize the matrix in Eq.~\eqref{eq:transformed_pde_1}, as follows:
\begin{align} \label{eq:A_matrix_1}
    \bm{A} &:= - \left( \frac{1}{2} \sum_{\mu=0}^{d-1} \left( D_\mu^{+} \tilde{\kappa} D_\mu^{-} + D_\mu^{-} \tilde{\kappa} D_\mu^{+} \right) + \sum_{\mu=0}^{d-1} \left( \tilde{\beta}_\mu^{+} D_\mu^{-} + \tilde{\beta}_\mu^{-} D_\mu^{+} \right) + \tilde{\alpha} \right) \nonumber \\
    &\quad + \frac{1}{2} \sum_{\mu=0}^{d-1} \chi_\mathrm{N}(B_\mu^{-}) D_\mu^{+} \tilde{\kappa} \left( I^{\otimes \sum_{\mu'=\mu+1}^{d-1}} \otimes \sigma_{00}^{\otimes n_\mu} \otimes I^{\otimes \sum_{\mu'=0}^{\mu-1} n_{\mu'}} \right) \nonumber \\
    &\quad - \frac{1}{2} \sum_{\mu=0}^{d-1} \chi_\mathrm{N}(B_\mu^{+}) D_\mu^{-} \tilde{\kappa} \left( I^{\otimes \sum_{\mu'=\mu+1}^{d-1}} \otimes \sigma_{11}^{\otimes n_\mu} \otimes I^{\otimes \sum_{\mu'=0}^{\mu-1} n_{\mu'}} \right) \nonumber \\
    &\quad + \frac{1}{2} \sum_{\mu=0}^{d-1} \chi_\mathrm{D}(B_\mu^{-}) \tilde{\kappa} \left( I^{\otimes \sum_{\mu'=\mu+1}^{d-1}} \otimes \sigma_{00}^{\otimes n_\mu} \otimes I^{\otimes \sum_{\mu'=0}^{\mu-1} n_{\mu'}} \right) \nonumber \\
    &\quad + \frac{1}{2} \sum_{\mu=0}^{d-1} \chi_\mathrm{D}(B_\mu^{+}) \tilde{\kappa} \left( I^{\otimes \sum_{\mu'=\mu+1}^{d-1}} \otimes \sigma_{11}^{\otimes n_\mu} \otimes I^{\otimes \sum_{\mu'=0}^{\mu-1} n_{\mu'}} \right),
\end{align}
where $\tilde{\beta}_\mu^{+} := \max(\tilde{\beta}_\mu, 0)$ and $\tilde{\beta}_\mu^{-} := \min(\tilde{\beta}_\mu, 0)$ for the upwind difference scheme~\cite{allaire2007numerical}, and the second and subsequent lines of Eq.~\eqref{eq:A_matrix_1} appear due to the boundary condition.
Note again that the discretization in Eq.~\eqref{eq:A_matrix_1} is just one example and we can use another finite difference scheme.
The ODE $\mathrm{d} \bm{w} / \mathrm{d} t = - \bm{A} \bm{w}$ under the initial condition $w_{j}(0) = u(0, \bm{x}^{[j]})$ yields the solution of $w_{j}(t) = u(t, \bm{x}^{[j]})$.
Since all qubit operators in the matrix $\bm{A}$ are represented by the tensor products of operators $\sigma_{01}$, $\sigma_{10}$, $\sigma_{00}$, $\sigma_{11}$, and $I$, we can calculate the multiplication of two operators in a symbolic manner based on some simple rules such as $\sigma_{01} \sigma_{10} = \sigma_{00}$, without any access to the full matrix.
As we obtain the qubit operator representation of the matrix $\bm{A}$, in the next subsection we consider how to efficiently represent the diagonal operators in $\bm{A}$, such as $\tilde{\kappa}$ and $\tilde{\alpha}$. 
As we will see later, we actually need an assumption that the spatially varying parameters are piecewise constant functions so that the resulting qubit operators could be represented efficiently.

\subsubsection{Efficient operator representation of spatially varying parameters} \label{sec:logic_compression}

Let us consider again a spatially varying parameter $c(\bm{x})$ discretized on the lattice which results in a diagonal operator $\tilde{c}$, as introduced in Eq.~\eqref{eq:c_diag}.
Now, we discuss the case where $c(\bm{x})$ takes the value either $c_0$ or $c_1$; this case will be discussed in an example later. 
Let $\mathcal{I}$ denote the index set where $\forall j \in \mathcal{I}, c(\bm{x}^{[j]}) = c_1$.
Then, we can represent $\tilde{c}$ as 
\begin{align}
    \label{eq:coef_x}
    \tilde{c} &= \sum_{j \in \mathcal{I}} \left( c_1 - c_0 \right) \ket{j} \! \bra{j} + c_0 I^{\otimes n} \nonumber \\
    &= \sum_{j \in \mathcal{I}} \left( c_1 - c_0 \right) \sigma_{j_{n-1} j_{n-1}} \otimes \dots \otimes \sigma_{j_0 j_0} + c_0 I^{\otimes n},
\end{align}
where $I$ is the identity operator, $(j_{n-1} \dots j_0)_2$ is the binary representation of $j$.
Thus, naively, we can represent $\tilde{c}$ using $| \mathcal{I} | + 1$ qubit operators.

We can reduce the number of terms by utilizing the relationship of $\sigma_{00}+\sigma_{11}=I$.
Specifically, we aim to perform factorization such as $\sigma_{jj} \otimes \sigma_{00} +\sigma_{jj} \otimes \sigma_{11} = \sigma_{jj} \otimes (\sigma_{00} +\sigma_{11}) = \sigma_{jj} \otimes I$. 
This example corresponds to the transformation of Boolean function $f = ab +a\bar{b} = a(b+\bar{b})=a$, where $a$ and $b$ are binary logic values.
The problem of minimizing the number of terms in Eq.~\eqref{eq:coef_x} corresponds to simplifying the Sum of Products (SOP) in Boolean algebra to represent a certain logic function with fewer terms. 

A variety of methodologies have been developed to tackle this type of problem, such as Quine-McCluskey algorithm~(QMC)~\cite{quine1952problem, mccluskey1956minimization} and Espresso algorithm~(EA)~\cite{brayton1982comparison, brayton1984logic}. 
QMC searches all candidate of the solution, hence it has the exact solution capabilities albeit with exponential computational complexity~\cite{chandra1978number}. 
Although some improved methods in scaling are developed~\cite{ducsa2015enhancing}, it is suitable for small-scale problems. 
On the other hand, EA efficiently search for candidate solutions by eliminating redundancy, hence it is particularly efficient for scaling. 
Although the analytical evaluation of the compression rates, the number of variables, and computational complexity are not clarified, the empirical method have demonstrated that highly accurate approximate solutions can be obtained with small computational cost, as evidenced in various practical results~\cite{Espresso-gpu2021}. 
In this study, we employed EA, considering the balance between execution time and optimality.

\begin{figure}[t]
    \centering
    \includegraphics[width=\textwidth]{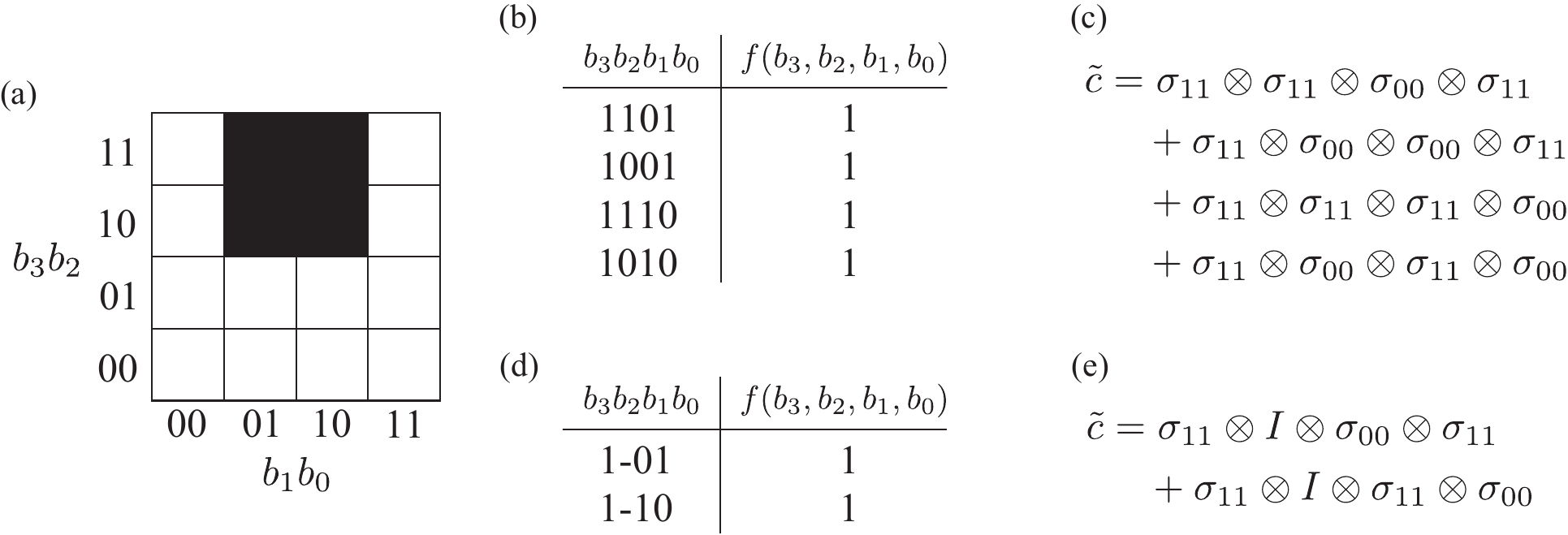}
    \caption{Concept of the simplification of operators for spatially varying parameters. (a) Example of spatially varying parameters on a $4 \times 4$ grid where black pixels have $c_1=1$ and white pixels have $c_0=0$. $x$ and $y$ coordinates corresponds to the bit strings $b_1 b_0$ and $b_3 b_2$, respectively.
    (b) Truth table of the Boolean function $f(b_3, b_2, b_1, b_0)$ which takes the value of $1$ when the corresponding pixel is black. 
    (c) Diagonal operator corresponding to the spatially varying parameters given in (a) before simplification. 
    (d) Truth table of the Boolean function $f(b_3, b_2, b_1, b_0)$ after simplifying (b) by the logic minimization technique. (e) Diagonal operator corresponding to the spatially varying parameters given in (a) after simplification.}
    \label{fig:lm_concept}
\end{figure}

Figure~\ref{fig:lm_concept} illustrates the conceptual diagram of simplification of the operators for spatially varying parameters by a logic minimization technique.
Let us consider the spatially varying parameters on a $4 \times 4$ grid where black pixels have the value of $c_1=1$ while white pixels have the value of $c_0=0$ as shown in Fig.~\ref{fig:lm_concept}(a). Associating the coordinate and the value of $c_1$ and $c_0$ with bit strings and binary values, respectively, we can write a truth table as shown in Fig.~\ref{fig:lm_concept}(b).
This truth table corresponds to the diagonal operator in Fig.~\ref{fig:lm_concept}(c), which is exactly Eq.~\eqref{eq:coef_x}.
Applying the logic minimization technique, specifically EA, for the Boolean function $f(b_3, b_2, b_1, b_0)$, we obtain the simplified Boolean function as shown in Fig.~\ref{fig:lm_concept}(d) where the symbol $-$ means that the output does not depend on the input variable, i.e., $b_2$.
This simplified truth table corresponds to the diagonal operator in Fig.~\ref{fig:lm_concept}(e), which is simplified from Fig.~\ref{fig:lm_concept}(c).
In this way, we can obtain the efficient (simplified) diagonal operator representation of spatially varying parameters by applying the EA for the truth table corresponding naive representation of spatially varying parameters.

Here, note the fact that our problem setting is not written by Boolean algebra, but each terms have continuous valued coefficients.
As far as Boolean algebra, the overlap of each terms is not an issue since the final output value is calculated by summing up all terms with the logical OR operation.
However, in our current problem setting, the overlap of different terms could cause the duplications of coefficient for a certain coordinate.
To avoid this, we have added a process to check for duplicate coordinates in the compressed logic obtained using EA. 
The specific workflow is described in Algorithm~\ref{alg:resolve_duplication}.
Note that this process for resolving duplicate coordinates can also be implemented efficiently, i.e., $O(|\mathcal{I}|^2 n)$.
It is also worth noting that the aforementioned procedure can be easily extended to the case where a spatially varying parameter $c(\bm{x})$ is a piecewise constant function.
That is, if the spatially varying parameters are piecewise constant functions, the logic minimizer technique is applicable for efficient representation of the operators.
Therefore, we assume in the present study that the spatially varying parameters are piecewise constant functions.
This assumption is reasonable in the engineering application because spatially varying parameters usually reflect the material properties in the computational domain, which makes spatially varying parameters be piecewise constant.
For instance, if we have two materials in the computational domain, spatially varying parameters take two values each of which corresponds to each material property.

\begin{algorithm}[t]
    \SetKwInOut{Input}{Input}
    \SetKwInOut{Output}{Output}
    \SetKwComment{Comment}{/* }{ */}

    \Input{List of bit strings \texttt{list\_orig}}
    \Output{List of modified bit strings \texttt{list\_mod}}
    \BlankLine

    Set \texttt{list\_mod} $\leftarrow ~ [\ ]$ \Comment*[r]{Initialize list as empty one.}
    
    \For{\texttt{b1} in \texttt{list\_orig}}{

        \For{\texttt{b2} in \texttt{list\_mod}}{

            \tcc{\texttt{get\_indices} returns indices of input bits which can take both 0 and 1.}
            \texttt{id1} $\leftarrow$ \texttt{get\_indices}(\texttt{b1})~;
            
            \texttt{id2} $\leftarrow$ \texttt{get\_indices}(\texttt{b2})~;
            
            \If{there is duplicate coordinate in \texttt{b1} and \texttt{b2}}{

                \For{\texttt{i} in \texttt{b1}}{

                    \If{\texttt{i} is not in \texttt{b2}}{

                        $\texttt{b1}[\texttt{i}] \leftarrow \texttt{b1}[\texttt{i}] \oplus 1$
                        \Comment*[r]{$\texttt{b1}[\texttt{i}]$ is the \texttt{i}th bit of \texttt{b1}. $\oplus$ is XOR.}
                    
                    }

                    \texttt{break}~;
                
                }    
            
            }
            
        }

        \texttt{b1} is appended to \texttt{list\_mod}~;
    
    }

 \caption{Resolving duplicate coordinate} \label{alg:resolve_duplication}
\end{algorithm}

\subsection{Coefficient oracle realized using matrix product state} \label{sec:coefficient_oracle}

Here, we discuss how to implement the LCU coefficient oracle $O_\mathrm{coef}$ in efficient and approximate manners via matrix product states (MPSs)~\cite{vidal2003efficient, schollwock2011density}. 
Suppose that the integer $a$ is written in $n_\mathrm{anc}$-bits binary representation as $a = (a_{n_\mathrm{anc}-1} \dots a_0)_2$, where $a_m \in \{0, 1\}$ is the $m$-th bit of the binary representation of $a$.
Here, we discretize the integration range of $k$ in Eq.~\eqref{eq:lchs_time_independent} into $2^{n_\mathrm{anc}}$ integration points $\{k_a \}_{a=0}^{2^{n_\mathrm{anc}}-1}$ ranging from $-2^{n_\mathrm{anc} - n_\mathrm{frac} - 1}$ to $2^{n_\mathrm{anc} - n_\mathrm{frac}} - 2$, at intervals of $2^{-n_\mathrm{frac}}$ where $n_\mathrm{frac} \in \mathbb{N}$. 
To this end, we define the integration point $k_a$, as follows:
\begin{align}
    k_{a= (a_{n_\mathrm{anc}-1} \dots a_0)_2} := \left( -a_{n_\mathrm{anc}-1} 2^{n_\mathrm{anc} - 1} + \sum_{m=0}^{n_\mathrm{anc} - 2} a_{m} 2^m \right) 2^{-n_\mathrm{frac}}.
\end{align}
This representation corresponds to associating $k_a$ with the binary $(a_{n_\mathrm{anc}-1} \dots a_0)_2$ regarded as the signed fractional binary with the integer part of $n_{\rm a} - n_{\rm frac} - 1$ bits and the fractional part of $n_{\rm frac}$ bits.
If an MPS corresponding to a quantum state $(1 / \sqrt{\| \bm{c} \|_1})\sum_{a=0}^{2^{n_\mathrm{anc}}-1} \sqrt{c_a} \ket{a}$ is obtained, a quantum circuit for preparing this quantum state can be constructed \cite{ran2020encoding, lubasch2020variational, rudolph2023decomposition, mc2024towards}.
Here, we set $\omega_a = 2^{-n_\mathrm{frac}}$, i.e., $c_a = 2^{-n_\mathrm{frac}} / \pi (1 + k_a^2)$ in the first step described in Section~\ref{sec:lchs}, which corresponds to the rectangular approximation of the integral with respect to $k$ in Eq.~\eqref{eq:lchs_time_independent}.
To prepare $(1 / \sqrt{\| \bm{c} \|_1})\sum_{a=0}^{2^{n_\mathrm{anc}}-1} \sqrt{c_a} \ket{a}$ in the form of an MPS, we first treat $[k_0 \dots k_{2^{n_\mathrm{anc}}-1}]^\top$ as a $2^{n_\mathrm{anc}}$-dimensional vector and represent it as an MPS as follows:
\begin{align} \label{eq:ka}
    \sum_{a=0}^{2^{n_\mathrm{anc}}-1} k_a \ket{a} = \sum_{a=0}^{2^{n_\mathrm{anc}}-1} \sum_{b_{n_\mathrm{anc}}=1}^{n_\mathrm{anc}} \dots \sum_{b_1=1}^{n_\mathrm{anc}} K_{a_{n_\mathrm{anc}} b_{n_\mathrm{anc}}}^{[n_\mathrm{anc}]} K_{b_{n_\mathrm{anc}} a_{n_\mathrm{anc} - 1} b_{n_\mathrm{anc} - 1}}^{[n_\mathrm{anc} - 1]} \dots K_{b_3 a_2 b_2}^{[2]} K_{b_2 a_1}^{[1]} \ket{a_{n_\mathrm{anc}} a_{n_\mathrm{anc} - 1} \dots a_1},
\end{align}
where
\begin{align}
    K_{a_{n_\mathrm{anc}} b_{n_\mathrm{anc}}}^{[n_\mathrm{anc}]} &= \begin{cases}
        - a_{n_\mathrm{anc}} 2^{n_\mathrm{anc} - 1 - n_\mathrm{frac}} & \text{ if } b_{n_\mathrm{anc}} = n_\mathrm{anc} \\
	1 & \text{otherwise}
    \end{cases} \label{eq:K_na} \\
    K_{b_{m+1} a_m b_m}^{[m]} &= \begin{cases}
        a_{m} 2^{m - 1 - n_\mathrm{frac}} & \text{ if } b_{m+1} = b_m = m \\
	\delta_{b_{m+1} b_{m}} & \text{otherwise}
    \end{cases} \label{eq:K_m} \\
    K_{b_2 a_1}^{[1]} &= \begin{cases}
	a_{1} 2^{-n_\mathrm{frac}} & \text{ if } b_2 = 1 \\
	1 & \text{otherwise}
    \end{cases}. \label{eq:K_1}
\end{align}
That is, $\sum_{a=0}^{2^{n_\mathrm{anc}}-1} k_a \ket{a}$ can be exactly represented as an MPS with the bond dimension of $n_\mathrm{anc}$.
Note that it is enough to store matrices in Eqs.~\eqref{eq:K_na}--\eqref{eq:K_1} to represent the MPS and we do not have to compute the exponentially many summation in Eq.~\eqref{eq:ka}.
Next, we derive an MPS that represents $\sum_{a=0}^{2^{n_\mathrm{anc}}-1} (1 + k_a^2) \ket{a}$ from an MPS representing $\sum_{a=0}^{2^{n_\mathrm{anc}}-1} k_a \ket{a}$ as follows:
\begin{align} \label{eq:mps_1_k_squared}
    \sum_{a=0}^{2^{n_\mathrm{anc}}-1} (1 + k_a^2) \ket{a} &= \sum_{a=0}^{2^{n_\mathrm{anc}}-1} \ket{a} + \sum_{a=0}^{2^{n_\mathrm{anc}}-1} k_a^2 \ket{a} \nonumber \\
	&= \sum_{a=0}^{2^{n_\mathrm{anc}}-1} \ket{a} \nonumber \\
	&\quad + \sum_{a=0}^{2^{n_\mathrm{anc}}-1} \sum_{b_{n_\mathrm{anc}}, \dots b_1=1}^{n_\mathrm{anc}} \sum_{b'_{n_\mathrm{anc}}, b'_1=1}^{n_\mathrm{anc}} K_{a_{n_\mathrm{anc}} b_{n_\mathrm{anc}}}^{[n_\mathrm{anc}]} \dots K_{b_2 a_1}^{[1]} K_{a_{n_\mathrm{anc}} b'_{n_\mathrm{anc}}}^{[n_\mathrm{anc}]} \dots K_{b'_2 a_1}^{[1]} \ket{a_{n_\mathrm{anc}} a_{n_\mathrm{anc} - 1} \dots a_1} \nonumber \\
	&= \sum_{a=0}^{2^{n_\mathrm{anc}}-1} \sum_{b_{n_\mathrm{anc}}, b_1=1}^{n_\mathrm{anc}^2+1} S_{a_{n_\mathrm{anc}} b_{n_\mathrm{anc}}}^{[n_\mathrm{anc}]} S_{b_{n_\mathrm{anc}} a_{n_\mathrm{anc} - 1} b_{n_\mathrm{anc} - 1}}^{[n_\mathrm{anc} - 1]} \dots S_{b_3 a_2 b_2}^{[2]} S_{b_2 a_1}^{[1]} \ket{a_{n_\mathrm{anc}} a_{n_\mathrm{anc} - 1} \dots a_1},
\end{align}
where
\begin{align}
    S_{a_{n_\mathrm{anc}} b_{n_\mathrm{anc}}}^{[n_\mathrm{anc}]} &= \begin{cases}
	K_{a_{n_\mathrm{anc}} \lfloor b_{n_\mathrm{anc}} / n_\mathrm{anc} \rfloor }^{[n_\mathrm{anc}]} K_{a_{n_\mathrm{anc}} \texttt{mod}(b_{n_\mathrm{anc}}, n_\mathrm{anc})}^{[n_\mathrm{anc}]} & \text{ if } b_{n_\mathrm{anc}} \leq n_\mathrm{anc}^2 \\
	1 & \text{otherwise}
    \end{cases} \label{eq:S_na} \\
    S_{b_{m+1} a_m b_m}^{[m]} &= \begin{cases}
	K_{\lfloor b_{m+1} / n_\mathrm{anc} \rfloor a_m \lfloor b_m / n_\mathrm{anc} \rfloor }^{[m]} K_{\texttt{mod}(b_{m+1}, n_\mathrm{anc}) a_m \texttt{mod}(b_m, n_\mathrm{anc})}^{[m]} & \text{ if } b_m \leq n_\mathrm{anc}^2 \\
        \delta_{b_{m+1} b_{m}} & \text{otherwise}
    \end{cases} \label{eq:S_m} \\
    S_{b_2 a_1}^{[1]} &= \begin{cases}
        K_{\lfloor b_2 / n_\mathrm{anc} \rfloor a_1 }^{[1]} K_{ \texttt{mod}(b_{2}, n_\mathrm{anc}) a_1}^{[1]} & \text{ if } b_2 \leq n_\mathrm{anc}^2 \\
	1 & \text{otherwise}
    \end{cases}. \label{eq:S_1}
\end{align}
That is, $\sum_{a=0}^{2^{n_\mathrm{anc}}-1} (1 + k_a^2) \ket{a}$ can also be exactly represented by an MPS with the bond dimension of $n_\mathrm{anc}^2+1$.
Note again that the calculation of Eq.~\eqref{eq:mps_1_k_squared} can be performed via calculation of each matrix in Eqs.~\eqref{eq:S_na}--\eqref{eq:S_1} and we do not have to compute the exponentially many summation in Eq.~\eqref{eq:mps_1_k_squared}.
However, we were unable to find an analytical method to represent the target $\sum_{a=0}^{2^{n_\mathrm{anc}}-1} \sqrt{1 + k_a^2} \ket{a}$ in an MPS form.
Therefore, we decided to obtain the target MPS through optimization, which utilizes $\sum_{a=0}^{2^{n_\mathrm{anc}}-1} (1 + k_a^2) \ket{a}$.
Consider the following vector-valued function:
\begin{align}
    &\mathcal{F}( \bm{\Psi} ) := \texttt{diag}(\bm{\Psi}) \bm{\Psi} -  \sum_{a=0}^{2^{n_\mathrm{anc}}-1} (1 + k_a^2) \ket{a},
\end{align}
where $\texttt{diag}$ is an operator that returns a diagonal matrix whose diagonal components corresponds to the input vector $\bm{\Psi}$.
Obviously, $\mathcal{F}( \bm{\Psi} )={\bf 0}$ iff $\bm{\Psi} = \sum_{a=0}^{2^{n_\mathrm{anc}}-1} \sqrt{1 + k_a^2} \ket{a}$, which is our target state.
In this study, we implement the Newton's method to find $\bm{\Psi}$ such that $\mathcal{F}( \bm{\Psi} )={\bf 0}$.
That is, using the Taylor expansion:
\begin{align}
    \mathcal{F}( \bm{\Psi} + \delta \bm{\Psi} ) &= \mathcal{F}( \bm{\Psi} ) + 2\texttt{diag}(\bm{\Psi}) \delta \bm{\Psi} + o(\| \delta \bm{\Psi} \|),
\end{align}
we start with an appropriate initial MPS vector $\ket{\Phi^0}$ and iteratively update by
\begin{align} \label{eq:mps_update}
    \bm{\Psi}^{[{n_\mathrm{it}}+1]} = \bm{\Psi}^{[n_\mathrm{it}]} + \delta \bm{\Psi}^{[n_\mathrm{it}]},
\end{align}
where ${n_\mathrm{it}}$ is the iteration count and $\delta \bm{\Psi}^{[n_\mathrm{it}]}$ is obtained by solving
\begin{align} \label{eq:mps_newton}
    2\texttt{diag}(\bm{\Psi}^{[n_\mathrm{it}]}) \delta \bm{\Psi}^{[n_\mathrm{it}]} = - \mathcal{F}( \bm{\Psi}^{[n_\mathrm{it}]} ).
\end{align}
Eq.~\eqref{eq:mps_newton} is a system of linear equations represented in MPS, and in this study, it was solved using the modified alternating least square (MALS) method~\cite{holtz2012alternating}.
Through this procedure, we can obtain an MPS that approximates $\sum_{a=0}^{2^{n_\mathrm{anc}}-1} \sqrt{1 + k_a^2} \ket{a}$ in the MPS form.
Finally, by solving the system of linear equations:
\begin{align} \label{eq:mps_inv}
    \texttt{diag}(\bm{\Psi}) \bm{\Phi} = \sqrt{\frac{2^{n_\mathrm{frac}}}{\pi}} \sum_{a=0}^{2^{n_\mathrm{anc}}-1} \ket{a},
\end{align}
we obtain the MPS $\bm{\Phi}$ that approximates the target state $\sum_{a=0}^{2^{n_\mathrm{anc}}-1} \sqrt{c_a} \ket{a}$ with $c_a = 2^{-n_\mathrm{frac}} / \pi (1 + k_a^2)$. 
We used the MALS method to solve Eq.~\eqref{eq:mps_inv}.  
Through these procedures, if the bond dimension of the MPS is kept at $\mathcal{O}(\mathrm{poly}(n_\mathrm{anc}))$, we can conduct a normalization of $\bm{\Phi}$ and construct an MPS that approximates $(1 / \sqrt{\| \bm{c} \|_1}) \sum_{a=0}^{2^{n_\mathrm{anc}}-1} \sqrt{c_a} \ket{a}$, within a time complexity of $\mathcal{O}(\mathrm{poly}(n_\mathrm{anc}))$.
Specifically, let $r_\Psi$ and $r_\Phi$ denote the bond dimensions for $\bm{\Psi}$ and $\bm{\Phi}$, respectively.
Since the time complexity of the MALS is $\mathcal{O}(\max(r_x, r_b)^3 r_A^2 n)$ for solving a linear system $Ax=b$ where $A$, $x$ and $b$ are MPO and MPSs with the bond dimensions of $r_A$, $r_x$ and $r_b$, respectively~\cite{holtz2012alternating}, the time complexity for solving Eq.~\eqref{eq:mps_newton} is $\mathcal{O}((r_\Psi^2 + n_\mathrm{anc}^2 + 1)^3 r_\Psi^2 n_\mathrm{anc})$. 
Also, the time complexity for solving Eq.~\eqref{eq:mps_inv} is $\mathcal{O}(r_\Phi^3 r_\Psi^2 n_\mathrm{anc})$.

\begin{figure}[t]
    \centering
    \includegraphics[width=\textwidth]{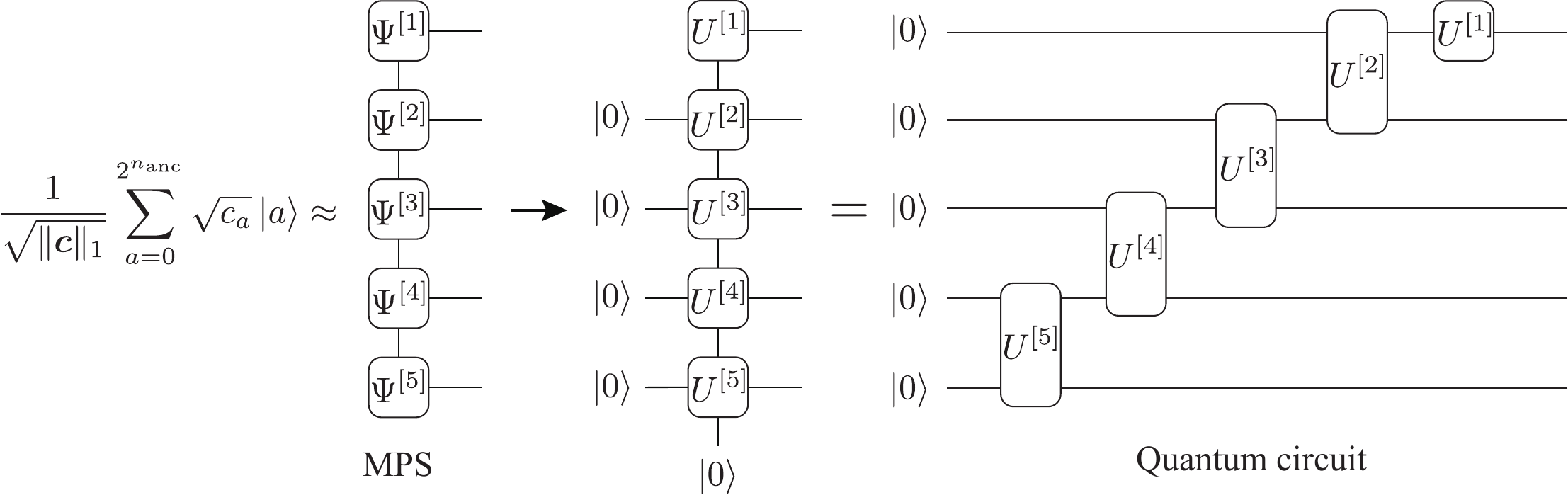}
    \caption{Constructing a quantum circuit from an MPS with the bond dimension of 2.}
    \label{fig:mps2circuit}
\end{figure}

The above procedures are conducted entirely through classical calculations of tensor networks. 
To realize this classical state on a quantum device, or in other words to obtain the coefficient oracle $O_\mathrm{coef}$, we use the scheme for converting an MPS into a quantum circuit~\cite{ran2020encoding, rudolph2023decomposition}. 
Figure~\ref{fig:mps2circuit} illustrates the procedure of constructing a quantum circuit from an MPS with the bond dimension of 2 for 5 qubits based on Ref.~\cite{ran2020encoding}.
Let $\bm{\Phi} \approx \sum_{a=0}^{2^{n_\mathrm{anc}}-1} \sqrt{c_a} \ket{a}$ with $c_a = 2^{-n_\mathrm{frac}} / \pi (1 + k_a^2)$ be a right canonical MPS.
For the rightmost matrix (topmost matrix in the figure), $\Psi^{[1]}$, which is unitary, we set $U^{[1]} = \Psi^{[1]}$.
For $1 < m < n_a = 5$, we set the $4 \times 4$ unitary $U^{[m]}$ as $U_{2a_m+b_m, 2b_{m+1}}^{[m]} = \Psi_{b_{m+1} a_m b_m}^{[m]}$ for even-numbered column and we set the odd-numbered column $U_{2a_m+b_m, 2b_{m+1}+1}^{[m]}$ so that $U^{[m]}$ can be unitary.
For the leftmost matrix, we set the $4 \times 4$ unitary $U^{[5]}$ as $U_{2a_5+b_5, 0}^{[m]} = \Psi_{a_5 b_5}^{[5]}$ for first column and we set other columns so that $U^{[5]}$ can be unitary.
Thus, we can construct a quantum circuit as shown in Fig.~\ref{fig:mps2circuit} from an MPS with the bond dimension of 2.
For MPSs with larger bond dimensions, we can also approximately construct the corresponding quantum circuits by repeatedly disentangling the MPS using an MPS with the bond dimension of 2~\cite{ran2020encoding,rudolph2023decomposition}.
Note that no access to the full state vector with the dimension of $2^{n_\mathrm{anc}}$ is necessary. 
The specific procedure is summarized in Algorithm~\ref{alg:coefficient_oracle}.

\begin{algorithm}[t]
    \SetKwInOut{Input}{Input}
    \SetKwInOut{Output}{Output}
    \SetKwComment{Comment}{/* }{ */}

    \Input{Bond dimensions $r_\Psi$ and $r_\Phi$, and tolerance $\epsilon$}
    \Output{Quantum circuit for $O_\mathrm{coef}$}
    \BlankLine

    Prepare an MPS for $\sum_{a=0}^{2^{n_\mathrm{anc}}-1} (1 + k_a^2)$ by Eq.~\eqref{eq:mps_1_k_squared};
    
    Set $n_\mathrm{it} \leftarrow 0$;
    
    Prepare a random MPS $\bm{\Psi}^{[0]}$ with the bond dimensions $r_\Psi$;

    \While{$\| \mathcal{F}(\bm{\Psi} \| > \epsilon $}{

        Solve Eq.~\eqref{eq:mps_newton} by MALS method for MPS with bond dimension $r_\Psi$;

        Update MPS by Eq.~\eqref{eq:mps_update};

        Set $n_\mathrm{it} \leftarrow n_\mathrm{it} + 1$;
        
    }

    Solve Eq.~\eqref{eq:mps_inv} by MALS  method for MPS with bond dimension $r_\Phi$;

    Use the MPS $\bm{\Phi} / \| \bm{\Phi} \|$ to construct the quantum circuit for oracle $O_\mathrm{coef}$ 

 \caption{Constructing coefficient oracle} \label{alg:coefficient_oracle}
\end{algorithm}

\subsection{Simulation algorithm and its complexity} \label{sec:procedure}

Here, we summarize the procedure for solving PDEs, which is a specific version of Section~\ref{sec:lchs}.
Because some parts of our proposed method include heuristic methods, it is difficult to derive the rigorous time complexity of our algorithm to simulate PDEs under a precision $\varepsilon$.
However, we herein provide the time complexity of each step depending on the predetermined hyper-parameters.

\begin{enumerate}
    \item Discretize the PDE of interest into an ODE as discussed in Section~\ref{sec:discretization}. 
    This includes the derivation of Hamiltonians $\bm{L}$ and $\bm{H}$ where spatially varying parameters are encoded using the logic minimization technique as described in Section~\ref{sec:logic_compression}.
    
    \item Prepare the system register consisting of $n$ qubits to embed the solution $\bm{w}(t)$ of the ODE and the ancilla register consisting of $n_\mathrm{anc}$ ancilla qubits for the LCU.
    The required number of integration points are $\mathcal{O}(\| \bm{L} \|_2 T / \varepsilon^2)$ where $T$ is a simulation time and $\varepsilon$ is a truncation error of integration~\cite{an2023linear}, which results in the number of ancilla qubits, $\mathcal{O}(\log (\| \bm{L} \|_2 T / \varepsilon^2))$.
    
    \item Construct the quantum circuit for the coefficient oracle $O_\mathrm{coef}$ on $n_\mathrm{anc}$-qubit ancilla system via the tensor network technique as described in Section~\ref{sec:coefficient_oracle}.
    The time complexity for solving Eq.~\eqref{eq:mps_newton} is $\mathcal{O}((r_\Psi^2 + n_\mathrm{anc}^2 + 1)^3 r_\Psi^2 n_\mathrm{anc})$ while the time complexity for solving Eq.~\eqref{eq:mps_inv} is $\mathcal{O}(r_\Phi^3 r_\Psi^2 n_\mathrm{anc})$.
    
    \item Initialize all the qubits to $\ket{0}^{\otimes n_\mathrm{anc} } \otimes \ket{0}^{\otimes n }$.
    
    \item Apply the state preparation oracle $O_\mathrm{prep}: \ket{0}^{\otimes n} \rightarrow \ket{w(0)} := \bm{w}(0) / \| \bm{w}(0) \|$ to the system register to prepare the quantum state $\ket{w(0)}$ representing the normalized initial state of $\bm{w}(0)$.
    We assume the analytic implementation for sparse distribution of $\bm{w}(0)$, or approximate implementation of the state preparation oracle~\cite{shirakawa2024automatic, nakaji2022approximate, melnikov2023quantum}.
    
    \item Apply the coefficient oracle for the LCU, $O_\mathrm{coef}: \ket{0} \rightarrow (1 / \sqrt{\| \bm{c} \|_1})\sum_{a=0}^{2^{n_\mathrm{anc}}-1} \sqrt{c_a} \ket{a}$, to the ancilla register.
        
    \item Alternatively apply the Hamiltonian simulation oracle $O_H$ to the system register and the select oracle $\texttt{SEL}_L$ to the whole system $r$ times, where $r := T / \tau$ is the number of steps for time evolution.
    Since the Hamiltonians $\bm{L}$ and $\bm{H}$ are represented by the operators $\sigma_{01}$, $\sigma_{10}$, $\sigma_{00}$, $\sigma_{11}$ and $I$, the time evolution operators of them, $O_L(\tau) := \exp(-i \bm{L} \tau)$ and $O_H(\tau) := \exp(-i \bm{H} \tau)$, can be implemented efficiently~\cite{sato2024hamiltonian}.
    The select oracle $\texttt{SEL}_L(\tau):= \sum_{a=0}^{2^{n_\mathrm{anc}}-1} \ketbra{a}{a} \otimes e^{-i k_a \bm{L} \tau}$ is implemented as  
    \begin{align}
        \texttt{SEL}_L(\tau) &:= \sum_{a=0}^{2^{n_\mathrm{anc}}-1} \ketbra{a}{a} \otimes e^{-i k_a \bm{L} \tau} \nonumber \\
        &= \sum_{a=0}^{2^{n_\mathrm{anc}}-1} \ketbra{a}{a} \otimes \exp \left(-i \left( -a_{n_\mathrm{anc}-1} 2^{n_\mathrm{anc} - 1} + \sum_{m=0}^{n_\mathrm{anc} - 2} a_{m} 2^m \right) 2^{-n_\mathrm{frac}} \bm{L} \tau \right) \nonumber \\
        &= \sum_{a=0}^{2^{n_\mathrm{anc}}-1} \ketbra{a}{a} \otimes \exp \left(i a_{n_\mathrm{anc}-1} 2^{n_\mathrm{anc} - 1 -n_\mathrm{frac}} \bm{L} \tau \right) \prod_{m=0}^{n_\mathrm{anc} - 2} \exp \left(-i a_{m} 2^{m - n_\mathrm{frac}} \bm{L} \tau \right) \nonumber \\
        &= \left( \ketbra{0}{0}_{n_\mathrm{anc}-1} \otimes I^{\otimes n} + \ketbra{1}{1}_{n_\mathrm{anc}-1} \otimes O_L(-2^{n_\mathrm{anc} - 1 -n_\mathrm{frac}} \tau) \right) \nonumber \\
        & \hspace{10mm} \prod_{m=0}^{n_\mathrm{anc} - 2} \left( \ketbra{0}{0}_m \otimes I^{\otimes n} + \ketbra{1}{1}_m \otimes 
        O_L(2^{m-n_\mathrm{frac}} \tau) \right), \tag{43}
    \end{align}
    where $\ketbra{0}{0}_m := I^{n_\mathrm{anc}-m-1} \otimes \ketbra{0}{0} \otimes I^{m}$ and $\ketbra{1}{1}_m := I^{n_\mathrm{anc}-m-1} \otimes \ketbra{1}{1} \otimes I^{m}$.
    That is, the select oracle $\texttt{SEL}_L(\tau)$ is implemented by Hamiltonian simulation $O_L$ controlled by each ancilla qubit as shown in Fig.~\ref{fig:overview}(d).
    Assuming that the number of terms of all diagonal operators for representing the spatially varying parameters is $\mathcal{O}(\text{poly}(n))$, the number of terms of $\bm{L}$ and $\bm{H}$ are $\mathcal{O}(\text{poly}(n))$, which results in quantum circuits for $O_H$ and $O_L$ with the depth of $\mathcal{O}(\text{poly}(n))$~\cite{sato2024hamiltonian}.
    
    \item Apply the inverse of the coefficient oracle, $O_\mathrm{coef}^\dagger$, to the ancilla register.
    This procedure prepares a quantum state approximating $(1 / \| \bm{c} \|_1 )\ket{0}^{\otimes n_\mathrm{anc}} \otimes \exp \left( - \bm{A} t \right) \ket{w(0)} + \ket{\perp}$ in the sense of numerical integration, where $\ket{\perp}$ represents a state orthogonal to the first term.
    
    \item Measure all qubits in the ancilla register. If the measurement outcome is all zeros, we obtain the system register that approximates the state in Eq.~\eqref{eq:lchs_time_independent} in the sense of numerical integration. 
    The expected number of repetition to successfully get measurement outcome of all zeros from the ancilla register is $\mathcal{O}(\| \bm{c} \|_1 \| \bm{w}(0) \| / \| \bm{w}(T) \|)$ using amplitude amplification where $\| \bm{c} \|_1 \sim \mathcal{O}(1)$~\cite{an2023linear}.
\end{enumerate}

\section{Result} \label{sec:result}
In this section, we provide several numerical experiments for demonstration.
We used Python and its library Qiskit 1.0~\cite{Qiskit}, the open-source toolkit for quantum computation, to implement quantum circuits.
We also used a Python library, Scikit-TT, for dealing with MPSs and the use of the MALS method~\cite{holtz2012alternating}.

\subsection{Acoustic simulation}

First, we perform a two-dimensional acoustic simulation in the system with a spatially varying speed of sound.
The governing equation for the velocity potential $\phi(t, \bm{x})$ is given as
\begin{align} \label{eq:acoustic}
    \frac{1}{c(\bm{x})^2} \pddif{\phi(t, \bm{x})}{t} = \nabla^2 \phi(t, \bm{x}), \text{ for } (t, \bm{x}) \in (0, T] \times \Omega, 
\end{align}
where $c(\bm{x})$ represents the speed of sound that can spatially vary. 
The velocity potential $\phi(t, \bm{x})$ leads to define the acoustic pressure to the ambient pressure $p(t, \bm{x})$, the velocity $\bm{v}(t, \bm{x})$, and the ambient density $\rho_0(\bm{x})$, as follows: 
\begin{align} \label{eq:potential_relationship}
    p(t, \bm{x}) = -\pdif{\phi(t, \bm{x})}{t}, ~ \bm{v}(t, \bm{x}) = \frac{1}{\rho_0(t, \bm{x})} \nabla \phi(t, \bm{x}).
\end{align}
See Appendix~\ref{sec:gov_acoustic} for the detailed derivation. 
The PDE~\eqref{eq:acoustic} corresponds to Eq.~\eqref{eq:pde} by setting $u(t, \bm{x}) = \phi(t, \bm{x})$, $\varrho(t, \bm{x}) = 1 / c(t, \bm{x})^2$, $\zeta(\bm{x}) = 0$, $\kappa(\bm{x}) = 1$, $\beta(\bm{x}) = 0$, $\alpha(\bm{x}) = 0$ and $f(\bm{x}) = 0$, which yield
\begin{align} \label{eq:transformed_acoustic}
    \frac{\partial }{\partial t} \begin{pmatrix}
        \frac{1}{c} \dot{\phi} \vspace{1mm} \\
        \nabla \phi 
    \end{pmatrix} &= - \begin{pmatrix}
        0 & -c \nabla^\top \vspace{1mm} \\
        -\nabla c & 0
    \end{pmatrix} \begin{pmatrix}
        \frac{1}{c} \dot{\phi} \vspace{1mm} \\
        \nabla \phi
    \end{pmatrix},
\end{align}
from Eq.~\eqref{eq:transformed_pde}.
Using the relationship of Eq.~\eqref{eq:potential_relationship}, Eq.~\eqref{eq:transformed_acoustic} reads
\begin{align} \label{eq:conservation}
    \frac{\partial }{\partial t} \begin{pmatrix}
        -\frac{1}{c} p \vspace{1mm} \\
        \rho_0 \bm{v}
    \end{pmatrix} &= - \begin{pmatrix}
        0 & -c \nabla^\top \vspace{1mm} \\
        -\nabla c & 0
    \end{pmatrix} \begin{pmatrix}
        -\frac{1}{c} p \vspace{1mm} \\
        \rho_0 \bm{v}
    \end{pmatrix},
\end{align}
which corresponds to the linearized conservation laws of mass and momentum with the constitution law of $p = c^2 \rho$.
Here, we impose the Dirichlet and Neumann boundary conditions for the left and right boundaries, respectively, which correspond to the sound soft and hard boundaries, respectively. 
Also, we impose the periodic boundary condition for the top and bottom boundaries.
These boundary conditions ensure Eq.~\eqref{eq:transformed_acoustic} to be the Schr\"{o}dinger equation with a Hermitian operator~\cite{sato2024hamiltonian}. 
Thus, we can simulate it by Eq.~\eqref{eq:lchs_time_independent} with $\bm{L}=0$, that is, the usual Hamiltonian simulation with unitary dynamics.
Specifically, the Hermitian and anti-Hermitian parts of the matrix $\bm{A}$ are
\begin{align}
    \bm{L} &= \frac{\bm{A} + \bm{A}^\dagger}{2} = 0 \nonumber \\
    \bm{H} &= \frac{\bm{A} - \bm{A}^\dagger}{2i} = i \sum_{\mu=0}^{d-1} \left( \ketbra{0}{\mu + 1} \otimes \tilde{c} D_\mu^{+} + \ketbra{\mu + 1}{0} \otimes D_\mu^{-} \tilde{c} \right).
\end{align}

\begin{figure}[t]
    \centering
    \includegraphics[width=0.3\textwidth]{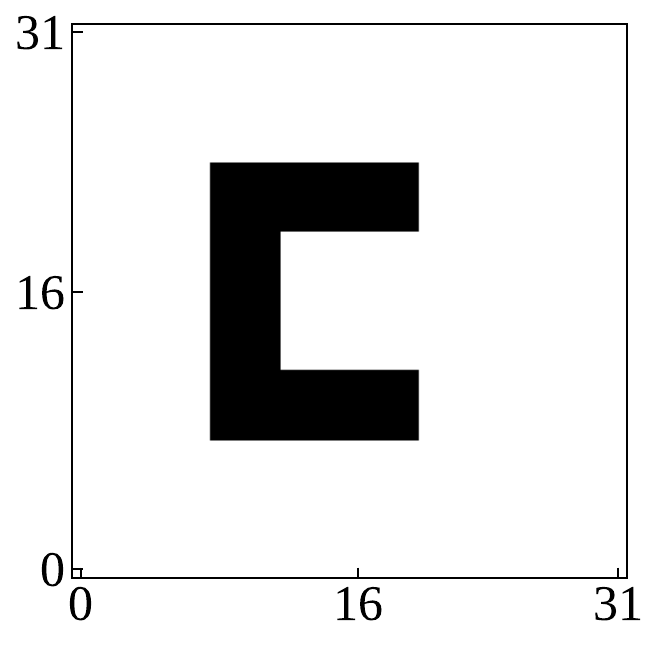}
    \caption{
    Distribution of the speed of sound $c(\bm{x})$ discretized on the lattice of $2^5 \times 2^5$ grid points. 
    In the black region $c(\bm{x})=10$, while $c(\bm{x})=1$ in the white region.
    }
    \label{fig:custom1}
\end{figure}
\begin{figure}[t]
    \centering
    \includegraphics[width=0.8\textwidth]{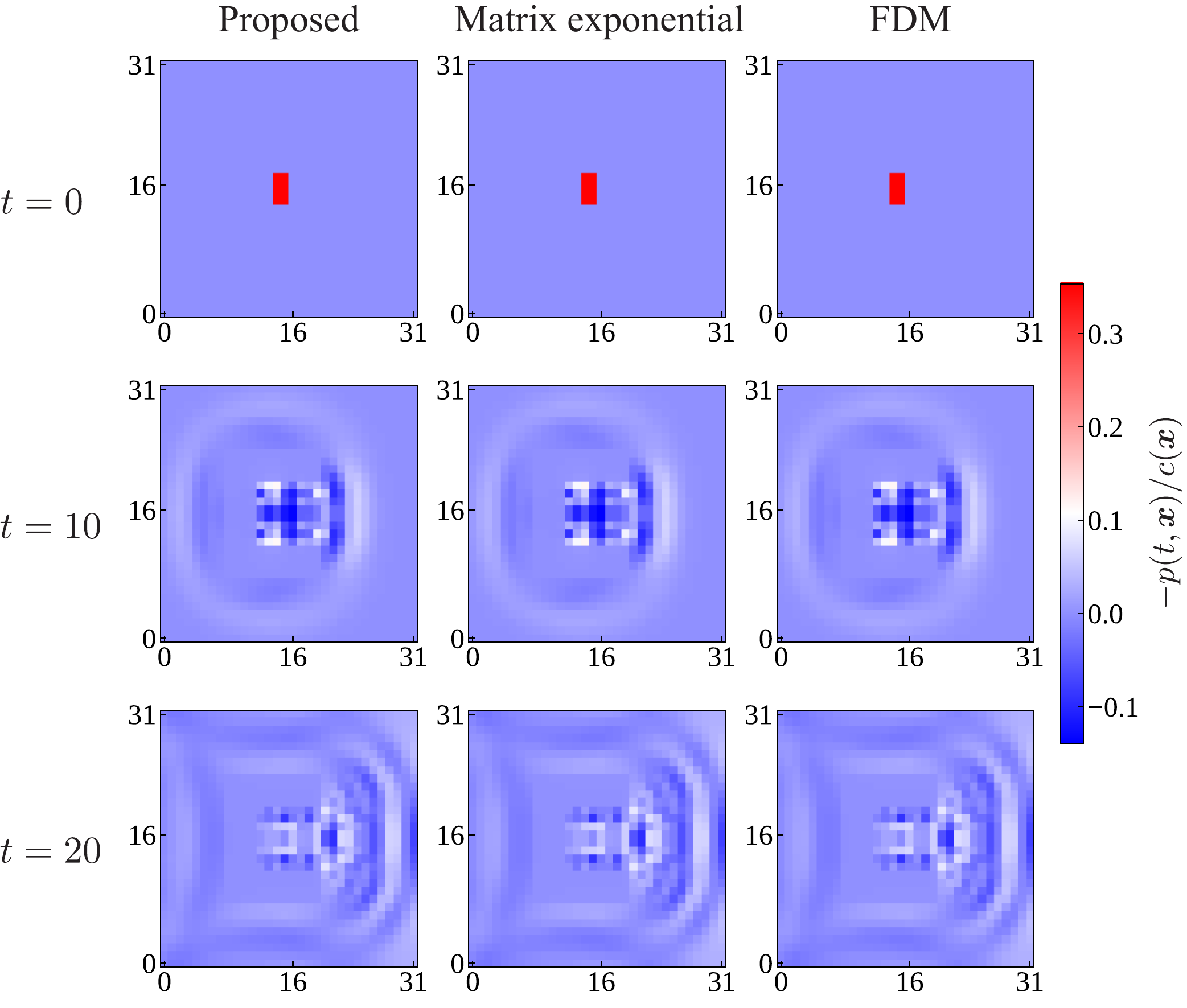}
    \caption{The contour of the field $\partial \phi(t, \bm{x}) / \partial t = -p(t, \bm{x}) / c(\bm{x})$ obtained by solving the acoustic equation.
    Each plot in the column represents the simulation results obtained by each approach.}
    \label{fig:result_wave_2d}
\end{figure}

In the simulation here we consider the speed of sound $c(\bm{x})$ is distributed as shown in Fig.~\ref{fig:custom1}: 
we prepare the lattice of $2^5 \times 2^5$ grid points, where $c(\bm{x})=10$ on the 128 nodes in the black region, while $c(\bm{x})=1$ in the white region. 
In this setting, the qubit operator for representing $c(\bm{x})$ naturally consists of $128+1$ terms in Eq.~\eqref{eq:coef_x}, but it successfully reduced to only six terms by using the logic minimizer technique described in Section~\ref{sec:logic_compression}.

To simulate the acoustic equation \eqref{eq:transformed_acoustic}, we set $n_0 = n_1 =5$, $\tau=1.0 \times 10^{-3}$, $T=20$ and $h=1$. 
Here we particularly simulate the quantum circuits by the state vector calculation.
The number of qubits for encoding this problem is $n_0 + n_1 + 2 = 12$, where the last two qubits are used for the first register of $\bm{w}(t)$ in Eq.~\eqref{eq:w_vector}.
As an initial state, we prepared
\begin{align}
    \bm{w}(0) = \ket{w(0)} = \frac{\sqrt{2}}{4} \ket{00} \otimes \left( \ket{01110} + \ket{01111} + \ket{10000} + \ket{10001} \right) \otimes \left( \ket{01110} + \ket{01111} \right),
\end{align}
which corresponds to the initial condition
\begin{align}
    &p(0, \bm{x}^{[j]}) = \begin{cases}
        -\frac{\sqrt{2}}{4}
        & \text{ for } x_1^{[j]}\in\{14,15,16,17\}, ~ x_0^{[j]} \in\{14,15\} \\
        0 & \text{ otherwise},
    \end{cases} \\
    &\bm{v}(0, \bm{x}^{[j]}) = 0.
\end{align}
This initial state can be generated by the state preparation oracle:
\begin{align}
    O_\mathrm{prep} = I^{\otimes 2} \otimes \underbrace{\mathrm{CX}_{10, 9} \mathrm{CX}_{10, 8} \mathrm{CX}_{10, 7} \left( H \otimes X^{\otimes 3} \otimes H \right)}_{\text{State preparation for } x_1^{[j]}} \otimes \underbrace{I \otimes X^{\otimes 3} \otimes H}_{\text{State preparation for } x_0^{[j]}},
\end{align}
where $\mathrm{CX}_{i, j}$ represents the CNOT gate acting on the $j$-th qubit controlled by the $i$-th qubit.
We herein use the little endian, i.e., we count the qubit from the right side.

We compare the simulation results of the proposed method with those obtained by directly using the matrix exponential operator $e^{-i \bm{H} \tau}$ to $\ket{w(0)}$ and those by the fully classical finite difference method (FDM) with the central differencing scheme.
We used the same time increment parameter $\tau=1.0 \times 10^{-3}$ for the FDM.

Figure~\ref{fig:result_wave_2d} illustrates the contour of the field $\partial \phi(t, \bm{x}) / \partial t = -p(t, \bm{x}) / c(\bm{x})$, which is a part of the elements of the state vector $\bm{w}(t)$, obtained by solving the acoustic equation. 
The left and center columns in Fig.~\ref{fig:result_wave_2d} represent solutions by the proposed quantum circuit and the matrix exponential operator, respectively.
The right column illustrates solutions calculated by the FDM.
Each row corresponds to the solutions at $t=0$, $t=10$ and $t=20$, respectively.
These results clearly coincide, implying that the proposed method could accurately simulate the acoustic equation with the spatially varying speed of sound shown in Fig.~\ref{fig:custom1}.
Actually, the dynamics of the acoustic pressure well describes how it is reflected at the interface of media as well as how it propagates.

Note that such direct visualization is impractical in a real quantum computation, because estimating all amplitudes of a quantum state requires an exponentially large number of measurements, which deteriorates the potential quantum advantage.
Thus, we have to set some observables to extract meaningful information from the quantum state prepared through the Hamiltonian simulation algorithm. 
One simple but useful observable for such purpose is the squared acoustic pressure, which is proportional to the acoustic intensity and can be extracted as 
\begin{align} \label{eq:intensity}
    \int_\Omega \chi_{\Omega_\mathrm{eval}}(\bm{x}) p^2 \mathrm{d} \bm{x} \approx \bm{w}^\dagger \left( \ketbra{0}{0} \otimes \tilde{c} \tilde{\chi} \tilde{c} \right) \bm{w}.
\end{align}
Here $\tilde{\chi}$ is the diagonal operator corresponding to the indicator function $\chi_{\Omega_\mathrm{eval}}(\bm{x})$ that takes $1$ if $\bm{x} \in \Omega_\mathrm{eval}$ and $0$ otherwise, where $\Omega_\mathrm{eval}$ is the evaluation domain.
Since the observable $\ketbra{0}{0} \otimes \tilde{c} \tilde{\chi} \tilde{c}$ is diagonal, the expectation value can be efficiently estimated by measurement outcomes of the computational basis.
In this way, if one is interested in a local quantity at a certain subdomain, one can evaluate such a quantity efficiently using the diagonal operator corresponding to the indicator function.
The power spectra~\cite{miyamoto2024quantum} is also a possible observable for an acoustic system.
We would like to investigate meaningful observables in more detail and also discuss efficient estimation of their expected values in our future work.

\subsection{Heat conduction simulation}

Next, we perform two-dimensional heat conduction simulation.
The governing equation for the temperature $\theta(t, \bm{x})$ is given as
\begin{align} \label{eq:heat}
    \pdif{\theta(t, \bm{x})}{t} = \kappa \nabla^2 \theta(t, \bm{x}), \text{ for } (t, \bm{x}) \in (0, T] \times \Omega,
\end{align}
which corresponds to Eq.~\eqref{eq:pde_1} by setting $u(t, \bm{x}) = \theta(t, \bm{x})$, $\kappa(\bm{x}) = \kappa$, $\beta(\bm{x}) = 0$, $\alpha(\bm{x}) = 0$ and $f(\bm{x}) = 0$.
Here, we impose the Dirichlet boundary conditions for all boundaries.
The Hermitian part $\bm{H}$ and the non-Hermitian part $\bm{L}$ 
of the matrix $\bm{A}$ are given by 
\begin{align}
    \bm{H} &= \frac{\bm{A} - \bm{A}^\dagger}{2i} = 0,
\nonumber \\
    \bm{L} &= \frac{\bm{A} + \bm{A}^\dagger}{2} = -\frac{\kappa}{2} \sum_{\mu=0}^{d-1} \left( D_\mu^{+} D_\mu^{-} + D_\mu^{-} D_\mu^{+} \right).
\end{align}
Thus, we use Eq.~\eqref{eq:lchs_time_independent} to simulate the heat equation as a non-unitary dynamics.

\begin{figure}[t]
    \centering
    \includegraphics[width=\textwidth]{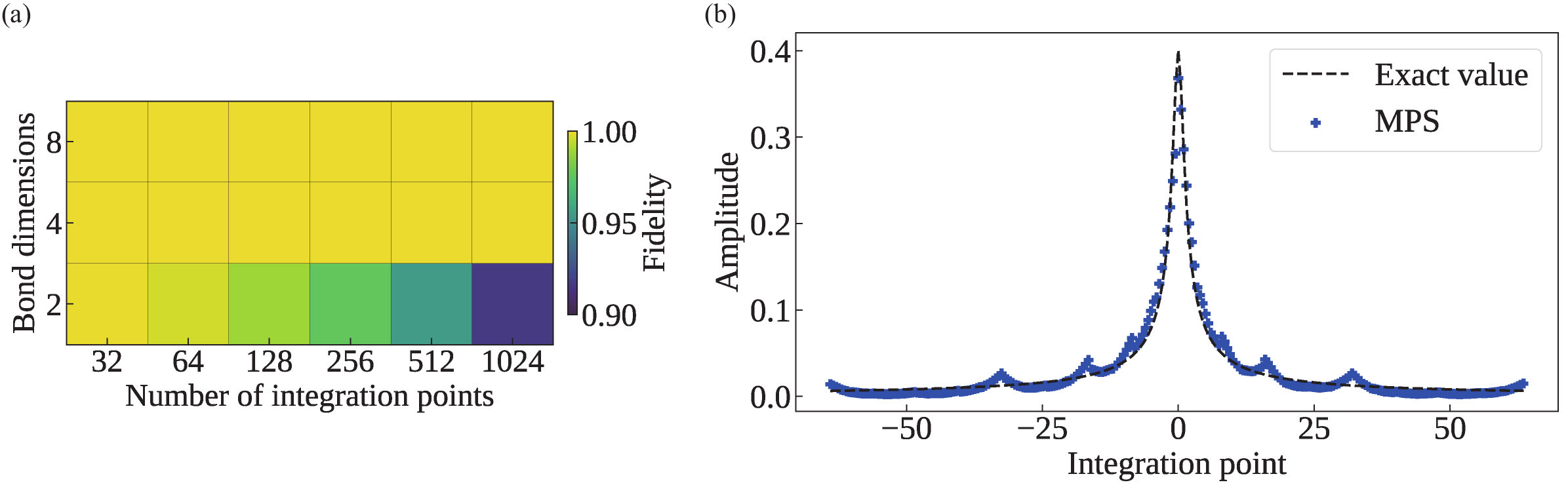}
    \caption{Result of preparing MPS for coefficient oracle $O_\mathrm{coef}$. (a) Heat map of fidelity between the exact state $(1 / \sqrt{\| \bm{c} \|_1})\sum_{a=0}^{2^{n_\mathrm{anc}}-1} \sqrt{c_a} \ket{a}$ and its approximation $\bm{\Phi}$ by MPS with a limited bond dimension $r_\Phi$ with respect to the number of integration points $2^{n_\mathrm{anc}}$. (b) The exact amplitude $\sqrt{2^{-n_\mathrm{frac}} / \pi (1 + k_a^2)}$ and its approximation by MPS with the bond dimensions $r_\Phi=2$ at each integration point $k_a$. }
    \label{fig:lcu_result}
\end{figure}
\begin{figure}[t]
    \centering
    \includegraphics[width=0.8\textwidth]{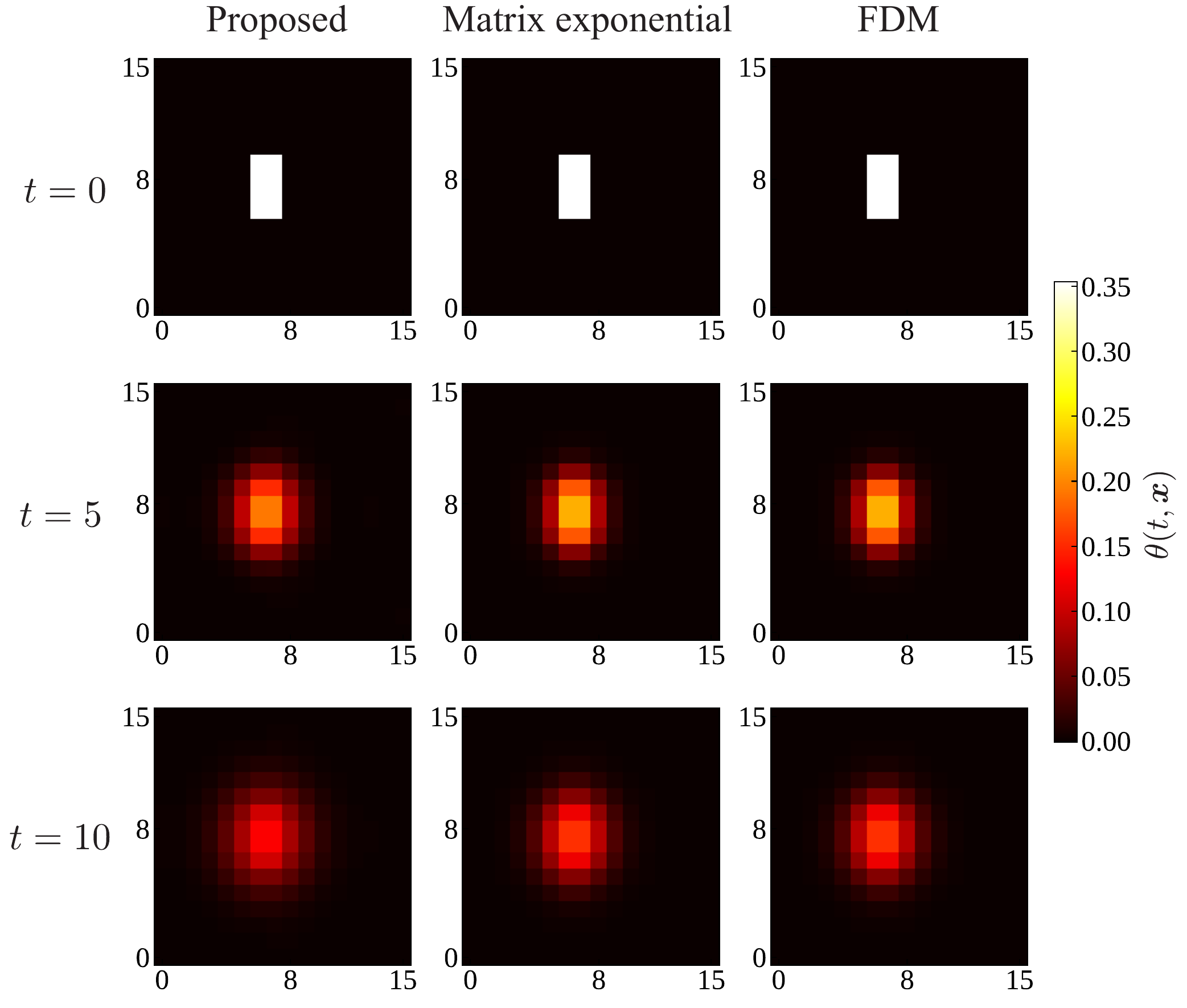}
    \caption{The contour of the field $\theta(t, \bm{x})$ obtained by solving the heat equation.
    Each plot in the column represents the simulation results obtained by each approach.}
    \label{fig:result_heat_2d}
\end{figure}

First, we examine the validity of the proposed MPS-based method (described in Section~\ref{sec:coefficient_oracle}) for implementing the LCU coefficient oracle $O_\mathrm{coef}$.
We set the number of bond dimensions for Eq.~\eqref{eq:mps_newton} as $r_\Psi=10$, the tolerance parameter as $\epsilon = 10^{-6}$, and the fraction part of integration points as $n_\mathrm{frac} = 1$.
Figure~\ref{fig:lcu_result} illustrates the result of the generated MPS for approximating $O_\mathrm{coef}$. 
Figure~\ref{fig:lcu_result}(a) compares the fidelity between the exact state $(1 / \sqrt{\| \bm{c} \|_1})\sum_{a=0}^{2^{n_\mathrm{anc}}-1} \sqrt{c_a} \ket{a}$ and the MPS $\bm{\Phi}$ with the bond dimension $r_\Phi$; the horizontal axis is the number of integration points $2^{n_\mathrm{anc}}$. 
This figure clearly shows that the bond dimension $r_\Phi \geq 4$ is large enough to well approximate the coefficient oracle with the fidelity greater than $0.999$.
The case of the bond dimension $r_\Phi = 2$ is also good for the relatively small number of the integration points, $2^{n_{\mathrm{anc}}} \leq 256$.
For example, the fidelity was $0.98$ when $2^{n_{\mathrm{anc}}} = 256$ and $r_\Phi = 2$.
Figure~\ref{fig:lcu_result}(b) shows the exact amplitude $\sqrt{2^{-n_\mathrm{frac}} / \pi (1 + k_a^2)}$ and its approximation by MPS for the case of $2^{n_{\mathrm{anc}}} = 256$ and $r_\Phi = 2$.
These results demonstrate the validity and the effectiveness of the proposed implementation method of the coefficient oracle $O_\mathrm{coef}$.
Since the coefficient oracle is not problem specific but is common for the implementation of LCHS, we confirm that our proposed method will work well for various applications based on LCHS.
In the following example, we set $n_{\mathrm{anc}} = 8$, $n_{\mathrm{frac}} = 1$, and $r_\Phi = 2$ because the MPS with the bond dimension of $2$ can be exactly implemented on a quantum circuit~\cite{ran2020encoding}.

To simulate the heat equation \eqref{eq:heat}, we set $n_0 = n_1 = 4$, $\tau=1.0 \times 10^{-1}$, $T=10$, $h=1$, and $\kappa=0.1$.
We simulated quantum circuits by the state vector calculation.
The number of qubits for encoding this problem is $n_0 + n_1 + n_\mathrm{anc} = 16$.
As an initial state, we prepared
\begin{align}
    \bm{w}(0) = \ket{w(0)} = \frac{\sqrt{2}}{4} \left( \ket{0110} + \ket{0111} + \ket{1000} + \ket{1001} \right) \otimes \left( \ket{0110} + \ket{0111} \right),
\end{align}
which corresponds to the initial condition
\begin{align}
    &\theta(0, \bm{x}^{[j]}) = \begin{cases}
        \frac{\sqrt{2}}{4}
        & \text{ for } x_1^{[j]}\in\{6,7,8,9\}, ~ x_0^{[j]} \in\{6,7\} \\
        0 & \text{ otherwise}.
    \end{cases}
\end{align} 
This initial state is generated using the following state preparation oracle:
\begin{align}
    O_\mathrm{prep} = \underbrace{\mathrm{CX}_{8, 7} \mathrm{CX}_{8, 6} \left( H \otimes X^{\otimes 2} \otimes H \right)}_{\text{State preparation for } x_1^{[j]}} \otimes \underbrace{I \otimes X^{\otimes 2} \otimes H}_{\text{State preparation for } x_0^{[j]}}.
\end{align}

We here compare the simulation results of the proposed method with those obtained by directly using the matrix exponential operator $e^{-i \bm{A} \tau}$ to $\ket{w(0)}$ and those computed by the fully classical finite difference method (FDM) with the central differencing scheme.
We use the time increment parameter $\tau=1.0 \times 10^{-3}$ for the FDM.

Figure~\ref{fig:result_heat_2d} illustrates the contour of the field $\theta(t, \bm{x})$, which is the solution of the heat equation.
The left and center columns in Fig.~\ref{fig:result_heat_2d} represent solutions by the proposed method and the matrix exponential operator, respectively.
The right column illustrates the solutions calculated by the FDM.
Each row corresponds to the solutions at $t=0$, $t=5$, and $t=10$, respectively.
These results well agree with each other, implying that the proposed method could simulate the heat equation by a quantum circuit, although the proposed method calculated the magnitude of the field $\theta(t, \bm{x})$ slightly smaller than those of other methods. 
The error between the left and center columns is attributed to the errors of Trotterization and numerical integration with respect to $k$, i.e., the implementation of coefficient oracle $O_\mathrm{coef}$.
Although the Trotterization error can be theoretically evaluated by the method in Refs.\cite{childs2021theory, layden2022first}, the implementation error of coefficient oracle $O_\mathrm{coef}$ is difficult to estimate due to its heuristic implementation via the tensor network.
Therefore, the bond dimension of the MPS would need to be sufficiently large to well approximate the exact state. 
The result shown in Fig.~\ref{fig:lcu_result}(a) provides a guideline, suggesting that, however, relatively low bond dimension is sufficient for such purpose. 
Reducing both errors occurred in Trotterization and the implementation of the coefficient oracle result in a longer quantum circuit. 
Therefore, it is essential to balance the circuit depth and the computational error for the choice of parameters.

This dynamics of the temperature describes how the initial temperature diffuses as time passes, which is a non-unitary dynamics.
Note again that such direct visualization requires the full quantum state tomography and thus is impractical for a real quantum computer. 
Furthermore, we need to repeatedly run quantum circuits to successfully measure all zeros of the ancilla register when implementing the proposed algorithm on the actual quantum computer. 
In this work, we performed all the computation by statevector simulatior to verify the validity of the proposed method.
We would like to conduct our future work that takes into account the sample complexity of required measurements for successfully extracting some meaningful information from the prepared time-evolved states.
When solving the heat equation, one might be interested in estimating the maximum temperature in the computational domain, which corresponds to the maximum amplitude of the time-evolved quantum state.
Although exactly finding the maximum amplitude is a difficult problem because of the exponentially large Hilbert space, it is easy to get an idea of whether the temperature is concentrated in a specific area or is more uniform throughout the domain by measuring multiple times in the computational basis. 
Then, we can also estimate the temperature of the identified regions using the amplitude estimation algorithm.
Also, the demonstration shown in the current work involved the two systems of PDE: one is a conservative with spatially varying parameters, and the other is non-conservative with uniform parameters, and did not tackle with non-conservative systems of PDEs with spatially varying parameters due to the computational cost although the formulation and algorithm of our proposed method include such systems.
We also address this in our future work.

\section{Conclusion} \label{sec:conclusion}

This paper proposed a quantum algorithm for solving linear partial differential equations (PDEs) of non-conservative systems with spatially varying parameters. 
To simulate the non-unitary dynamics, we employed the linear combination of Hamiltonian simulation (LCHS) method~\cite{an2023linear}, which is the Hamiltonian simulation version of the linear combination of unitaries (LCU). 
The key techniques for deriving quantum circuits of LCHS particularly for solving PDEs are (1) efficient encoding of heterogeneous media into the Hamiltonian by a logic minimization technique and (2) efficient quantum circuit implementation of LCU via the tensor-network technique.
Together with these key techniques, we provided a concrete recipe for constructing a quantum circuit for solving a target PDE. 
For demonstrating the validity of the proposed method, we focused on two PDEs, the acoustic and heat equations.
We employed the acoustic equation as an example of handling the spatially varying parameters, while we selected the heat equation as a typical example of a non-conservative system.
In numerical experiments, we confirmed that the simulation results of our proposed method agreed well with those obtained by the fully classical finite difference method, implying the wide implementability of PDEs on quantum circuits explicitly.

In the present study, we focused on PDEs without source terms (i.e., $\bm{b}(t)$ in Eq.~\eqref{eq:ode}) for simplicity, while the original LCHS method can deal with equations including the source term, as shown in Eq.~\eqref{eq:lchs}.
Since PDEs with source terms are essential for practical applications in CAE, we plan to address this problem in our future work. 
Also, we would like to investigate important observables in more detail for a specific problem setting.
For example, in addition to the local quantities at a specific region discussed in Eq.~\eqref{eq:intensity}, global (averaged) quantities such as effective stiffness or conductivity of a composite materials can be promising options in material science.
Additionally, our future work will tackle the simulation of non-linear PDEs, which have already been discussed theoretically in the literature~\cite{joseph2020koopman, tanaka2023quantum}. 
We believe our current work will advance the applicability of quantum computers in the field of CAE.

\section*{Acknowledgment}
This work is supported by MEXT Quantum Leap Flagship Program Grant Number JPMXS0118067285 and JP- MXS0120319794, and JSPS KAKENHI Grant Number 20H05966.

\appendix
\section{Mapping PDEs with the second-order derivative in time into those with the first-order derivative in time} \label{sec:mapping_pde}
To convert the second-order PDE~\eqref{eq:pde} in time into the first-order one, we define the vector field $\bm{w}(t, \bm{x})$, as follows:
\begin{align}
    \bm{w}(t, \bm{x}) = w_0(\bm{x}) \pdif{u(t, \bm{x})}{t} \ket{0} + \sum_{\mu=0}^{d-1} w_{\mu+1}(\bm{x}) \pdif{u(t, \bm{x})}{x_\mu} \ket{\mu + 1} + w_{d+1}(\bm{x}) u(t, \bm{x}) \ket{d+1},
\end{align}
where $\{ w_{\mu'}(\bm{x}) \}_{\mu'=0}^{d+1}$ are coefficient functions.
Since the success probability of solving PDE by LCU depends on the norm of the vector at the final time, we would like to determine the coefficient functions $\{ w_{\mu'}(\bm{x}) \}_{\mu'=0}^{d+1}$ so that the norm of the vector field $\bm{w}(t, \bm{x})$ is kept as much as possible.
The squared norm of the vector field is given as
\begin{align}
    &\int_\Omega \bm{w}(t, \bm{x})^\dagger \bm{w}(t, \bm{x}) \mathrm{d} \bm{x} \nonumber \\
    &= \int_\Omega \left( |w_0(\bm{x})|^2 \left| \pdif{u(t, \bm{x})}{t} \right|^2 + \sum_{\mu=0}^{d-1} |w_{\mu+1}(\bm{x})|^2 \left| \pdif{u(t, \bm{x})}{x_\mu} \right|^2 + |w_{d+1}(\bm{x})|^2 \left| u(t, \bm{x}) \right|^2 \right) \mathrm{d} \bm{x},
\end{align}
and its time derivative reads
\begin{align}
    &\int_\Omega \frac{\partial \bm{w}(t, \bm{x})^\dagger \bm{w}(t, \bm{x})}{\partial t} \mathrm{d} \bm{x} \nonumber \\
    &= \int_\Omega |w_0(\bm{x})|^2 \pdif{u^\ast(t, \bm{x})}{t} \pddif{u(t, \bm{x})}{t} \mathrm{d} \bm{x} + \sum_{\mu=0}^{d-1} \int_\Omega |w_{\mu+1}(\bm{x})|^2 \pdif{u^\ast(t, \bm{x})}{x_\mu} \frac{\partial^2 u(t, \bm{x})}{\partial t \partial x_\mu} \mathrm{d} \bm{x} \nonumber \\
    &\quad + \int_\Omega |w_{d+1}(\bm{x})|^2 u^\ast(t, \bm{x}) \pdif{u(t, \bm{x})}{t} \mathrm{d} \bm{x} + c.c. \nonumber \\
    &= -\int_\Omega |w_0(\bm{x})|^2 \frac{\zeta(\bm{x})}{\varrho(\bm{x})} \left| \pdif{u(t, \bm{x})}{t} \right|^2 \mathrm{d} \bm{x} + \sum_{\mu=0}^{d-1} \int_\Omega \frac{|w_0(\bm{x})|^2}{\varrho(\bm{x})} \pdif{u^\ast(t, \bm{x})}{t}  \pdif{}{x_\mu} \left( \kappa(\bm{x}) \pdif{u(t, \bm{x})}{x_\mu} \right) \mathrm{d} \bm{x} \nonumber \\
    &\quad  - \int_\Omega \frac{|w_0(\bm{x})|^2}{\varrho(\bm{x})} \pdif{u^\ast(t, \bm{x})}{t} \alpha(\bm{x}) u(t, \bm{x}) \mathrm{d} \bm{x} \nonumber \\
    &\quad + \sum_{\mu=1}^{d-1} \int_\Omega  |w_{\mu+1}(\bm{x})|^2 \pdif{u^\ast(t, \bm{x})}{x_\mu} \frac{\partial^2 u(t, \bm{x})}{\partial t \partial x_\mu} \mathrm{d} \bm{x} + \int_\Omega |w_{d+1}(\bm{x})|^2 u^\ast(t, \bm{x}) \pdif{u(t, \bm{x})}{t} \mathrm{d} \bm{x} + c.c. \nonumber \\
    &= -\int_\Omega |w_0(\bm{x})|^2 \frac{\zeta(\bm{x})}{\varrho(\bm{x})} \left| \pdif{u(t, \bm{x})}{t} \right|^2 \mathrm{d} \bm{x} + \sum_{\mu=0}^{d-1} \int_\Omega \frac{|w_0(\bm{x})|^2}{\rho(\bm{x})} \pdif{u^\ast(t, \bm{x})}{t}  \pdif{}{x_\mu} \left( \kappa(\bm{x}) \pdif{u(t, \bm{x})}{x_\mu} \right) \mathrm{d} \bm{x} \nonumber \\
    &\quad + \sum_{\mu=0}^{d-1} \int_{\partial \Omega} \frac{\partial u^\ast(t, \bm{x})}{\partial t} |w_{\mu+1}(\bm{x})|^2 \pdif{u(t, \bm{x})}{x_\mu} n_\mu(\bm{x}) \mathrm{d} \bm{x} - \sum_{\mu=0}^{d-1} \int_{\Omega} \pdif{u^\ast(t, \bm{x})}{t} \frac{\partial}{\partial x_\mu} \left( |w_{\mu+1}(\bm{x})|^2 \pdif{u(t, \bm{x})}{x_\mu} \right) \mathrm{d} \bm{x} \nonumber \\
    &\quad - \int_\Omega \frac{|w_0(\bm{x})|^2}{\varrho(\bm{x})} \pdif{u^\ast(t, \bm{x})}{t} \alpha(\bm{x}) u(t, \bm{x}) \mathrm{d} \bm{x} + \int_\Omega \pdif{u^\ast(t, \bm{x})}{t} |w_{d+1}(\bm{x})|^2 u(t, \bm{x}) \mathrm{d} \bm{x} + c.c. \nonumber \\
    &=-\int_\Omega |w_0(\bm{x})|^2 \frac{\zeta(\bm{x})}{\varrho(\bm{x})} \left| \pdif{u(t, \bm{x})}{t} \right|^2 \mathrm{d} \bm{x} \nonumber \\
    &\quad + \sum_{\mu=0}^{d-1} \int_\Omega \pdif{u^\ast(t, \bm{x})}{t} \left( \frac{|w_0(\bm{x})|^2}{\varrho(\bm{x})} \pdif{}{x_\mu} \left( \kappa(\bm{x}) \pdif{u(t, \bm{x})}{x_\mu} \right) - \pdif{}{x_\mu} \left( |w_{\mu+1}(\bm{x})|^2 \pdif{u(t, \bm{x})}{x_\mu} \right) \right)  \mathrm{d} \bm{x} \nonumber \\
    &\quad + \sum_{\mu=0}^{d-1} \int_{\partial \Omega} \frac{\partial u^\ast(t, \bm{x})}{\partial t} |w_{\mu+1}(\bm{x})|^2 \pdif{u(t, \bm{x})}{x_\mu} n_\mu(\bm{x}) \mathrm{d} \bm{x} \nonumber \\
    &\quad + \int_\Omega \pdif{u^\ast(t, \bm{x})}{t} \left( -\frac{|w_0(\bm{x})|^2}{\varrho(\bm{x})} \alpha(\bm{x}) + |w_{d+1}(\bm{x})|^2 \right) u(t, \bm{x}) \mathrm{d} \bm{x} + c.c.,
\end{align}
where we used the PDE in Eq.~\eqref{eq:pde} for the second equality and used the integration by parts for the third equality.
By setting $w_0(\bm{x}) = \sqrt{\varrho(\bm{x})}$, $w_{\mu+1}(\bm{x}) = \sqrt{\kappa(\bm{x})}$ and $w_{d+1}(\bm{x}) = \sqrt{\alpha(\bm{x})}$, most terms vanishes and the time derivative of the squared norm ends up
\begin{align}
    \int_\Omega \frac{\partial \bm{w}(t, \bm{x})^\dagger \bm{w}(t, \bm{x})}{\partial t} \mathrm{d} \bm{x} &= - 2 \int_\Omega \zeta(\bm{x}) \left| \pdif{u(t, \bm{x})}{t} \right|^2 \mathrm{d} \bm{x} \leq 0, 
\end{align}
where the term integrating over the boundary $\partial \Omega$ vanishes owing to the boundary condition.
This equation implies that the norm of the vector field $\bm{w}(t, \bm{x})$ monotonically decreases during the time evolution by the PDE~\eqref{eq:pde} depending on the damping factor $\zeta(\bm{x})$.
That is, the damping factor $\zeta(\bm{x})$ dominates the decrease of the norm and therefore the success probability of LCU while other coefficients have no effect on the norm.

\section{Changes in the norm by PDEs with the first-order derivative in time} \label{sec:norm_pde_1}
Since the success probability of solving PDEs by LCU depends on the changes in the norm over the simulation time, we would like to evaluate the changes in the norm by PDEs with the first-order derivative in time as well.
The time derivative of the squared norm of the field $u(t, \bm{x})$ is calculated as
\begin{align}
    & \int_\Omega \frac{\partial u(t, \bm{x})^\dagger u(t, \bm{x})}{\partial t} \mathrm{d} \bm{x} \nonumber \\
    & = \int_\Omega u(t, \bm{x})^\dagger \left( \nabla \cdot \kappa \nabla  - \bm{\beta} \cdot \nabla - \alpha \right) u(t, \bm{x}) \mathrm{d} \bm{x} + c.c \nonumber \\
    & = \int_{\partial \Omega} u(t, \bm{x})^\dagger \bm{n} \cdot \kappa \nabla u(t, \bm{x}) \mathrm{d} \bm{x} - \int_\Omega \left( \nabla u(t, \bm{x})\right)^\dagger \cdot \kappa \nabla u(t, \bm{x}) \mathrm{d} \bm{x} \nonumber \\
    & \quad - \int_\Omega u(t, \bm{x})^\dagger \bm{\beta} \cdot \nabla u(t, \bm{x}) \mathrm{d} \bm{x} - \int_\Omega \alpha u(t, \bm{x})^\dagger u(t, \bm{x}) \mathrm{d} \bm{x} + c.c. \nonumber \\
    & = -2 \int_\Omega \left( \nabla u(t, \bm{x})\right)^\dagger \cdot \kappa \nabla u(t, \bm{x}) \mathrm{d} \bm{x} -2 \int_\Omega \alpha \left| u(t, \bm{x}) \right|^2 \mathrm{d} \bm{x} \nonumber \\
    & \quad -\int_{\partial \Omega} \left| u(t, \bm{x}) \right|^2 \bm{n} \cdot \bm{\beta} \mathrm{d} \bm{x} + \int_\Omega \left| u(t, \bm{x}) \right|^2 \nabla \cdot  \bm{\beta} \mathrm{d} \bm{x},
\end{align}
where the first term in the third line integrating over the boundary $\partial \Omega$ vanishes owing to the boundary condition.
In the last equality, we applied the integration by parts for the complex conjugate term of the third term, which is about the advection.
This equation implies that the norm of the field $u(t, \bm{x})$ monotonically decreases over the time evolution by the PDE~\eqref{eq:pde_1} when the velocity field $\bm{\beta}(\bm{x})$ satisfies the condition that the last two terms vanish, i.e., the conservative law $\nabla \cdot \bm{\beta}(\bm{x}) = 0$ and $\bm{n} \cdot \bm{\beta}(\bm{x}) = 0$.
When such a condition is satisfied, the diffusion and absorption coefficients $\kappa(\bm{x})$ and $\alpha(\bm{x})$ determine the decrease of the norm and consequently the success probability of LCU.

\section{Acoustic equation} \label{sec:gov_acoustic}
To derive the governing equation for acoustic simulation, we consider the conservation laws of mass and momentum given as
\begin{align}
    \pdif{\rho(t, \bm{x})}{t} + \nabla \cdot \left( \rho(t, \bm{x}) \bm{v}(t, \bm{x}) \right) &= 0 \nonumber \\
    \rho \pdif{ \bm{v}(t, \bm{x})}{t} + \rho(t, \bm{x}) \left( \bm{v}(t, \bm{x}) \cdot \nabla \right) \bm{v}(t, \bm{x}) &= - \nabla p(t, \bm{x})
\end{align}
where $\rho(t, \bm{x})$ is the density, $\bm{v}$ is the velocity, and $p(t, \bm{x})$ is the pressure.
By regarding acoustic disturbances as small perturbations to an ambient state, we obtain the linear approximation of the conservation laws~\cite{pierce2019acoustics}, as follows:
\begin{align}
    \pdif{\rho'(t, \bm{x})}{t} + \nabla \cdot \left( \rho_0(\bm{x}) \bm{v}'(t, \bm{x}) \right) &= 0 \nonumber \\
    \rho_0(\bm{x}) \pdif{ \bm{v}'(t, \bm{x})}{t} &= - \nabla p'(t, \bm{x}),
\end{align}
with the relationship of $\rho(t, \bm{x}) = \rho_0(\bm{x}) + \rho'(t, \bm{x})$, $p(t, \bm{x}) = p_0 + p'(t, \bm{x})$, $\bm{v}(t, \bm{x}) = \bm{v}'(t, \bm{x})$ where $\rho_0$ and $p_0$ are ambient density and pressure, and $\rho'$, $\bm{v}'$ and $p'$ are small perturbations to the ambient states.
The density and pressure satisfies the following constitution law:
\begin{align}
    p'(t, \bm{x}) &= c(\bm{x})^2 \rho'(t, \bm{x}),
\end{align}
where $c(\bm{x})$ is the spatially varying sound speed.
Introducing the velocity potential $\phi(t, \bm{x})$ such that $\rho_0(\bm{x}) \bm{v}'(t, \bm{x}) = \nabla \phi(t, \bm{x})$ and considering the conservation and constitution laws, we obtain the governing equation for the potential, as follows:
\begin{align}
    \frac{1}{c(\bm{x})^2} \pddif{\phi(t, \bm{x})}{t} = \nabla^2 \phi(t, \bm{x}),
\end{align}
where the following relationships hold:
\begin{align}
    p'(t, \bm{x}) = \pdif{\phi(t, \bm{x})}{t}, ~ \bm{v}'(t, \bm{x}) = \frac{1}{\rho_0(\bm{x})} \nabla \phi(t, \bm{x}).
\end{align}
Note that the velocity potential is usually introduced such that $\bm{v}'(t, \bm{x}) = \nabla \phi(t, \bm{x})$~\cite{pierce2019acoustics}.
However, we define it as above to take into account the spatial varying density $\rho_0(\bm{x})$.

\bibliography{ref}

\end{document}